\documentclass[12pt, fleqn]{article}
\usepackage[cp1251]{inputenc}
\usepackage{latexsym,amsfonts,amssymb}
\usepackage{graphicx}

\usepackage{amsbsy}
\usepackage{amsmath}
\usepackage{epsf}
\usepackage{cite}

\newtheorem{remark}{Remark}
\newcommand{\lf}{\left}
\newcommand{\rg}{\right}

\newcommand{\be} {\begin{equation}}
\newcommand{\ee} {\end{equation}}
\newcommand{\ba} {\begin{array}{l}}
\newcommand{\ea} {\end{array}}
\newcommand{\p} {\partial}

\begin{document}

\begin{center}
{\Large \bf Reaction-diffusion equations \\ in mathematical models
arising in epidemiology }

\medskip

{\bf Vasyl' Davydovych$^{a}$, Vasyl' Dutka$^{b}$ and Roman Cherniha
$^{a,c,}$\footnote{\small Corresponding author. E-mail:
r.m.cherniha@gmail.com; roman.cherniha1@nottingham.ac.uk} }

$^{a}$ \quad Institute of Mathematics,  National Academy of Sciences
of Ukraine, \\
 3, Tereshchenkivs'ka Street, Kyiv 01004, Ukraine\\
$^{b}$ \quad Bakul Institute for Superhard Materials, National Academy of Sciences of Ukraine, \\
2, Avtozavods'ka Street, Kyiv 04074, Ukraine\\
 $^{c}$ \quad School of Mathematical Sciences, University of Nottingham,\\
  University Park, Nottingham NG7 2RD, UK
\end{center}

 \begin{abstract}
The review is devoted to analysis  of mathematical models used for
describing epidemic processes.  A main focus is done  on the models
that are based on partial differential equations (PDEs), especially
those that  were developed and used for  the COVID-19 pandemic
modelling. Our attention is paid preferable to the  studies in which
not only results of numerical simulations are presented but
analytical results as well.  In particular, travelling fronts
(waves), exact
 solutions, estimation of key epidemic parameters  of
  the epidemic  models with governing PDEs (typically
reaction-diffusion equations) are discussed. The  review  may serve
as a valuable source  for researchers and practitioners in the field
of mathematical  modelling in epidemiology.
\end{abstract}


\section{Introduction} \label{sec-1}

At the present time, there are numerous mathematical models
describing epidemic processes that  can be found in several  books
dedicated to mathematical modelling in life sciences (see
\cite{brauer-12,diekmann-2000,keeling-2008,murray-1989,murray-2003,hadeler-2017,bailey-1975}
and papers cited therein). It is widely thought that the
Kermack--McKendrick  study \cite{kermack-1927} is a pioneering paper
in this direction. The authors created a model based on three
ordinary differential equations (ODEs).  Nowadays their model  is
called the Susceptible-Infectious-Recovered (SIR) model. There are
several generalizations of the SIR model such as the
Susceptible-Exposed-Infectious-Recovered (SEIR) model suggested  in
\cite{dietz,anderson} and Susceptible-Infectious-Recovered-Deceased
(SIRD) model \cite{kermack-1932,bailey-1975,lin-2010}. Some other
models  (see, e.g., \cite{ch-dav-preprint20,ch-dav-2020,yang-2021})
were developed after the outbreak of the COVID-19 coronavirus
because this novel pandemic  has attracted extensive attention of
many mathematicians working in the field of mathematical modelling.

The classical pandemic models, first of all  SIR, SEIR and SIRD,
have been extensively used in the modelling of the COVID-19 pandemic
in order to understand and predict the transmission dynamics of the
disease (see, e.g.,
\cite{nesteruk-2021,fanelli-2020,alenezi-2021,jai-2022,kalachev-2023,kyrychko-2020}).
In particular, these models and their variants can be used to
estimate key epidemiological parameters of COVID-19, such as the
basic reproduction number $R_0$ (see, e.g.,
\cite{alimohamadi-2020,alenezi-2021,salom-2021}). The number $R_0$
represents an average number of new infections caused by a single
infected individual in the susceptible population.

Typically, numerical simulations are used for solving the above
mentioned models \cite{barlow-2020,chen-2021,zhu-2021}. It should be
noted that numerical solving of ODEs is not a cutting edge problem
at the present time because there are many computer program packages
adopted for these purposes. However, exact solving of the well-known
models (SIR, SEIR, SIRD, etc.) is a nontrivial problem because the
relevant ODEs are nonlinear. There are several studies focused on
investigating and constructing exact solutions of the classical SIR
model, its generalizations, modifications, simplification (see,
\cite{nucci-2004,harko-2014,ch-dav-preprint20,ch-dav-2020,
yoshida-2022,yoshida-2023} and papers cited therein). There are also
studies focused on solving such ODE systems using approximate
techniques. For instance,  the SIR epidemic model is solved by the
homotopy analysis method and  solutions are derived in the form of
series involving exponents in \cite{khan-2009}. In
\cite{marinca-2021}, the SEIR model is studied and approximate
solutions are obtained using a so-called optimal auxiliary functions
method.

It is worth noting that the propagation of numerous epidemic
processes, including the COVID-19 pandemic, often exhibits
significant spatial heterogeneity. For example, an essential spatial
heterogeneity was observed in large European countries (Italy is a
typical example) and  USA during the first pandemic wave. This
observation can be interpreted in various manners, but the
prevailing approach involves partitioning the larger spatial domain
(such as a country) into multiple smaller sub-domains (regional
divisions) and employing standard ODE-based models to each
sub-domain. Nevertheless, an alternative method exists -- using the
reaction-diffusion equations
 -- to model the diffusion-based spread of the infected population.
 Reaction-diffusion systems are used to
describe the dynamics of spatially extended systems in which the
interactions between components involve both reaction and diffusion
processes. These models have been widely applied in various fields,
including physics, chemistry, biology, and epidemiology, in order to
understand the spread and behavior of populations or substances.
When considering epidemic modelling, reaction-diffusion systems can
be used to study the spatial spread of infectious diseases (see
pioneering
\cite{kendall-1965,radcliffe-1973,noble-1974,murray-1985,kallen-1984}
and  recent works \cite{zhang-2020,zhi-2023,viguerie-2020,
grave-2021,viguerie-2021,grave-2022,tu-2023,mammeri-2020,cheng-2022,zhu-2019,yin-2022}).

There are also epidemic models that involve more complicated
equations, such as models based on reaction-diffusion systems with
cross-diffusion
\cite{radcliffe-1973,capasso-1994,bendahmane-2010,ch-dav-2022-cd,bailey-1975}.
It should be noted that cross-diffusion phenomena occurs in other
biological processes and was introduced in 1970s, in particular, for
mathematical  modelling   in  chemotaxis \cite{kel-seg-71} and
population dynamics \cite{sh-ka-te}. Nowadays such mathematical
models are intensively studied by different mathematical techniques
(see, e.g., \cite{ch-da-ki-23} and references therein).

Age-structured epidemic models should be mentioned as well. Such
models based on  integro-differential equations  in order to
incorporate both spatial and age dimensions for modelling  the
spread of infectious diseases (see
\cite{kuniya-2015,kuniya-2018,kang-2021,tian-2022} and papers cited
therein). These
 models aim to
understand the complex dynamics of disease transmission in relation
to age-specific factors and spatial patterns.

This review is organized as follows. In Section~\ref{sec-2}, we
briefly present classical models (SIR, SEIR, etc.) describing
epidemic processes with the stress on integrability of such models.
In Section~\ref{sec-3}, we present epidemic  models based on systems
of reaction-diffusion equations. Some of them were developed before
the COVID-19 pandemic, others are suggested very recently. In
Section~\ref{sec-4}, models involving PDEs with convective terms and
those with cross-diffusion are discussed. In Section~\ref{sec-5},
age-structured epidemic models are considered. Finally, we present
conclusions and briefly  discuss  unsolved problems in epidemic
modelling in the last section.

\section{Integrability of the classical epidemic models}\label{sec-2}

As mentioned above, the   SIR model is  a pioneering model for
describing epidemic processes. The model is bases on the
three-component system of ODEs that reads as  \cite{kermack-1927}
 \be\label{2-1}\ba
\frac{dS}{dt}=-\alpha SI, \medskip \\
\frac{dI}{dt}=\alpha SI-\beta I, \medskip \\
\frac{dR}{dt}=\beta I.
 \ea\ee
 The model divides the total population into three subpopulations
  (compartments): susceptible
($S$), infectious ($I$) (with infectious capacity and not yet
recovered) and recovered ($R$) (recovered and not be either
infectious or infected once again). In (\ref{2-1}), $\alpha>0$ is
the transmission rate (represents the average rate at which
susceptible individuals become infected when they come into contact
with infectious individuals); $\beta>0$ is the recovery rate
(represents the average rate at which infectious individuals recover
from the disease and gain immunity). The main assumption of the SIR
model consists in conservation law of  total population.
Mathematically it directly follows from (\ref{2-1}) if one takes the
sum of all equations
\be\label{2-1**}\frac{dS}{dt}+\frac{dI}{dt}+\frac{dR}{dt}=0, \ee
i.e. the total population $N(t)= S(t)+I(t)+R(t)=N_0$ ($N_0$ is a
fixed number). It means that the epidemic process does not lead to
deaths, i.e. the death rate is zero (each  member of inflicted
subpopulation will survive). Obviously, this assumption is rather
unrealistic for such processes as  the COVID-19 pandemic. It should
be also noted that the SIR model neglects the natural birth/death
rate.

The general solution of the nonlinear ODE system (\ref{2-1}) cannot
be expressed explicitly, but it is well-known in parametric form
(see, e.g., \cite{harko-2014}):
 \be\nonumber
S(t)=S_0\tau, \
I(t)=\frac{\beta}{\alpha}\ln\tau-S_0\tau-\frac{I_0}{\alpha}, \
R(t)=-\frac{\beta}{\alpha}\ln\tau, \ee
 where $S_0$ and $I_0$ are  integration constants, while the parameter
  $\tau$ is defined by the integral
 \be\label{2-2*} \int\frac{d\tau}{\tau\lf(I_0-\beta\ln\tau+\alpha S_0\tau\rg)}=t.
 \ee
The above integral leads to special functions, therefore a
transcendent functional equation is obtained for finding the
parameter $\tau$. Obviously,  $S_0$ and $I_0$  allow us to satisfy
initial conditions for the functions $S(t)$ and $I(t)$ (the third
constant $R_0$ springs up from the above integral).

In paper \cite{harko-2014}, the authors also consider the SIR model
with birth and death rates (see (4)--(6) \cite{harko-2014})  and
show that the relevant nonlinear system of ODEs can be reduced to
the Abel equation, which  is not solvable. However,  the authors
studied the equation obtained using semianalytical/numerical
methods.

The most known  generalization of the SIR model that assumes nonzero
death rate for the infected subpopulation is the SIRD model. The
SIRD model takes into account the following assumptions:
there is no cure or immunity and some  infected members of the
subpopulation $I$ will die at a certain rate;
the recovered subpopulation will not remain immune and can be again
infected (in contrast  to the  SIR model where the recovered group
obtains immunity from the disease).
Consequently, the SIRD model is described by the following system of
differential equations (see, e.g., \cite{lin-2010}):
 \be\label{2-8}\ba
\frac{dS}{dt}=-\alpha SI, \medskip \\
\frac{dI}{dt}=\alpha SI-\beta I-\mu I, \medskip \\
\frac{dR}{dt}=\beta I, \medskip \\
\frac{dD}{dt}=\mu I,
 \ea\ee
where $D$ represents the number of individuals who have died due to
the disease;  new parameter $\mu$  is the death rate.

In \cite{yoshida-2022}, an exact solution of the  SIRD epidemic
model (\ref{2-8}) is constructed
 for arbitrary
  initial conditions
\[S(0)=S_0, \ I(0)=I_0, \ R(0)=R_0, \ D(0)=D_0, \quad S_0+I_0+R_0+D_0=N_0\]
in the parametric form \be\label{2-9}\ba
S(t)=S_0\exp\lf(\frac{\alpha}{\beta}\,R_0\rg)\tau, \medskip\\
I(t)=N_0-D_0+\frac{\mu}{\beta}R_0-S_0\exp\lf(\frac{\alpha}{\beta}\,R_0\rg)\tau
+\frac{\beta+\mu}{\alpha}\ln\tau,
\medskip\\ R(t)=-\frac{\beta}{\alpha}\ln\tau, \medskip\\
D(t)= D_0-\frac{\mu}{\alpha}\ln\tau-\frac{\mu}{\beta}R_0, \ea\ee
where the parameter
$\tau\in\lf(\exp\lf(-\frac{\alpha}{\beta}\,R(\infty)\rg),
\exp\lf(-\frac{\alpha}{\beta}\,R_0\rg)\rg],$
 \[t=\int_{\tau}^{\exp\lf(-\frac{\alpha}{\beta}\,R_0\rg)}\frac{dz}{z\psi(z)},\]
while $\psi(z)=\alpha(N_0-D_0)+\frac{\alpha\mu}{\beta}\,R_0-\beta
S_0\exp\lf(\frac{\alpha}{\beta}\,R_0\rg)z+(\beta+\mu)\ln z.$
Notably, the above formula for the parameter $\tau$  is similar to
(\ref{2-2*}).

One notes that the sum of all components in (\ref{2-9}) is  $N_0$,
i.e. the conservation law again preserved. However, the total
population in this case (in contrast to the SIR model) includes
those who died from the epidemic disease.

The natural generalization of the SIR model is the SEIR one
\be\label{2-3}\ba
\frac{dS}{dt}=-\alpha SI, \medskip \\
\frac{dE}{dt}=\alpha SI-\gamma E, \medskip \\
\frac{dI}{dt}=\gamma E-\beta I, \medskip \\
\frac{dR}{dt}=\beta I.
 \ea\ee
 Model (\ref{2-3}) includes a new function $E(t)$  for the exposed
 subpopulation, representing
  individuals who have been exposed to the infection but are currently in a latent period, not yet capable of transmitting the disease.
In (\ref{2-3}), the coefficient $\gamma>0$ denotes the transition
rate from exposed individuals to the infected one and determines an
incubation period $\frac{1}{\gamma}$ that represents the average
duration from the time of infection to the point at which an
individual becomes capable of transmitting the disease to others.

Clearly, the SEIR model is developed also under assumption that the
conservation law of total population takes place. One may say that
the SEIR model is a straightforward generalization of the SIR model.

In recent work \cite{yoshida-2023}, an exact solution of the SEIR
model (\ref{2-3}) with the initial conditions \[S(0)=S_0, \
E(0)=E_0, \ I(0)=I_0, \ R(0)=R_0, \quad S_0+E_0+I_0+R_0=N_0\] is
constructed. Similarly to the SIR model case, the exact solution is
found  in the  parametric form:
 \be\label{2-4}\ba
S(t)=S_0\exp\lf(\frac{\alpha}{\beta}\,R_0\rg)\tau, \medskip\\
E(t)=E_0\exp\lf(-\gamma\varphi(\tau)\rg)+S_0\exp\lf(\frac{\alpha}{\beta}\,R_0-
\gamma\varphi(\tau)\rg)\int_{\tau}^{\exp\lf(-\frac{\alpha}{\beta}\,R_0\rg)}\exp\lf(\gamma\varphi(z)\rg)dz, \medskip\\
I(t)=N_0-S_0\exp\lf(\frac{\alpha}{\beta}\,R_0\rg)\tau+\frac{\beta}{\alpha}\ln\tau-E_0\exp\lf(-\gamma\varphi(\tau)\rg)-\\
\hskip2cm S_0\exp\lf(\frac{\alpha}{\beta}\,R_0-
\gamma\varphi(\tau)\rg)\int_{\tau}^{\exp\lf(-\frac{\alpha}{\beta}\,R_0\rg)}\exp\lf(\gamma\varphi(z)\rg)dz,
\medskip\\ R(t)=-\frac{\beta}{\alpha}\ln\tau, \ea\ee where the parameter $\tau\in\lf(\exp\lf(-\frac{\alpha}{\beta}\,R(\infty)\rg),
\exp\lf(-\frac{\alpha}{\beta}\,R_0\rg)\rg],$
 \[t=\varphi(\tau)=\int_{\tau}^{\exp\lf(-\frac{\alpha}{\beta}\,R_0\rg)}\frac{dz}{z\psi(z)}.\]
Here the function $\psi(z)$ is the solution of the Abel  equation of
the second kind
\[z\psi\psi'-(\beta+\gamma)\psi+\beta\gamma\ln z-\alpha\gamma\exp\lf(\frac{\alpha}{\beta}\,R_0
\rg)z+\alpha\gamma N_0=0, \] on the interval
\[\lf(\exp\lf(-\frac{\alpha}{\beta}\,R(\infty)\rg),
\exp\lf(-\frac{\alpha}{\beta}\,R_0\rg)\rg), \] that satisfies the
boundary conditions  \be\nonumber\ba
\psi\lf(\exp\lf(-\frac{\alpha}{\beta}\,R_0\rg)\rg)=\beta
I_0,\\
\lim\limits_{z\rightarrow\exp\lf(-\frac{\alpha}{\beta}\,R(\infty)\rg)+0}\psi(z)=0.\ea\ee
Moreover, a natural requirement about positivity of the solution
should be fulfilled \[ \psi(z)>0 \ \texttt{in} \
\lf(\exp\lf(-\frac{\alpha}{\beta}\,R(\infty)\rg),
\exp\lf(-\frac{\alpha}{\beta}\,R_0\rg)\rg].\] An applicability of
the exact  solution (\ref{2-4})  for   practical applications is
questionable because that is too cumbersome.

The natural  generalization of the SEIR model (\ref{2-3}) that
assumes nonzero death rate for infected subpopulation is the
 Susceptible-Exposed-Infected-Recovered-Deceased (SEIRD)
model
 \be\nonumber\ba
\frac{dS}{dt}=-\alpha SI, \medskip \\
\frac{dE}{dt}=\alpha SI-\gamma E, \medskip \\
\frac{dI}{dt}=\gamma E-\beta I-\mu I, \medskip \\
\frac{dR}{dt}=\beta I, \medskip \\
\frac{dD}{dt}=\mu I,
 \ea\ee
where all parameters have the same interpretations as  above. To the
best of our knowledge,  there are no papers devoted to the search
for exact solutions of the SEIRD type models.
 However, these models were used in several  papers (see, e.g.,
\cite{zhu-2021,loli-2020}) for modelling the COVID-19 pandemic.
Typically, results of  numerical  simulations are presented in such
papers that   describe  the dynamics of the pandemic and to suggest
its control strategies.

There are some other models describing epidemic processes that are
based on ODEs systems, however, those cannot be considered as direct
generalizations of the SIR model. Interestingly that some of them
are integrable and here we present examples.

In \cite{nucci-2004}, the authors investigate the model
 \be\label{2-5}\ba
\frac{dS}{dt}=-\alpha SI - \mu S + \beta I + \mu K, \medskip \\
\frac{dI}{dt}=\alpha SI-(\mu+\beta)I,
 \ea\ee
 that can be thought as a simplification of the classical SIR model.
 However, it is rather difficult to identify assumptions that reduce
 the SIR model to the ODE system (\ref{2-5}).
The authors assume  that the parameter $\mu$ is the proportionate
death rate, while the term  $\mu K$ represents a constant birth
rate.  Integrability of this model is proved  by means of the
Painlev\'{e} analysis. Moreover, using the Lie symmetry analysis,
the model was completely  integrated. As a result,
 the following exact solution in terms of elementary
functions was constructed \be\nonumber\ba
S(t)=\frac{\mu+\beta}{\alpha}+I(t)\big(\mu+\beta-\alpha K+I(t)-\mu
Be^{-\mu t}\big), \medskip\\
I(t)=\frac{1}{\alpha}\frac{A\exp\lf[-(\mu+\beta -\alpha K)t +
Be^{-\mu t} \rg]}{A\mu\int\exp\lf[-(\mu+\beta -\alpha K)t + Be^{-\mu
t} \rg]dt+C}.\ea\ee Here $A, \ B$ and $C$ are arbitrary constants.
However, it can be noted that one of them can be skipped because
three constants cannot vanish simultaneously.
 Two remaining constants can be used in order to satisfy initial conditions.

In \cite{ch-dav-preprint20, ch-dav-2020}, we proposed   the model
 \begin{equation}\label{2-10}\ba \frac{du}{dt} =  u(a-bu^{\kappa}),\medskip \\
\frac{dv}{dt} = k(t)u, \medskip \\ u(0)=u_0\geq0, \ v(0)=v_0\geq0,
  \ea\end{equation}
for quantitative description of the outbreak of the COVID-19
pandemic. In (\ref{2-10}), a smooth function $u(t)$ presents the
total number of the COVID-19 cases identified up to day (the time
moment) $t$; $ v(t)$ is the total number of deaths
 up to the time moment $t$. Typically, $t$ is an  integer number  but we assume that $u(t)$ and $ v(t)$ are continuous functions similarly to the functions $S, E,  I, R$ and $D$ used above.
One may also note that the relations between $u(t)$ and $ v(t)$:
\[ u(t)=\int^t_0 I(\tau)d\tau, \quad v(t)=\int^t_0 D(\tau)d\tau, \]
assuming that the pandemic started at the moment $t=0$.

In (\ref{2-10}), $a>0$ is the coefficient for the virus transmission
mechanism; $b>0$ is the coefficient for the effectiveness of the
government restrictions (quarantine rules); $\kappa>0$ is the
exponent, which guarantees that the total number of the COVID-19
cases is bounded in time; the smooth function $k(t)>0$ is the
coefficient for effectiveness of the health care system during the
epidemic process. From mathematical point of view, coefficient
$k(t)$ should have the asymptotic behavior $k(t)\rightarrow 0$, if
$t\rightarrow\infty$, otherwise all infected people will die. It is
assumed that $k(t)=k_0\exp(-\alpha t), \  \alpha> 0$. Note that in
the case $\kappa=1$ the first equation of model (\ref{2-10})
coincides with the classical logistic equation \cite{verh-1838} that
occurs naturally in epidemiology as  it was shown under some general
assumptions in \cite{brauer-12}.

The general  solution of   model (\ref{2-10}) is constructed
explicitly in the form
\begin{equation}\label{2-11} \begin{array}{l} \medskip  u(t)=a^{1/\kappa}u_0 e^{a t}\Bigg(a+b\,u_0^{\kappa}(e^{a \kappa t}-1)\Bigg)^{-1/\kappa},
 \\ v(t)= a^{1/\kappa}k_0u_0\int^t_0e^{(a-\alpha)\tau}\Bigg(a+b\,u_0^{\kappa}(e^{a \kappa
 \tau}-1)\Bigg)^{-1/\kappa}\,d\tau+v_0.
 \end{array}\end{equation}

The integral in (\ref{2-11}) can be expressed via  special functions
for arbitrary parameters $a$, $\alpha$ and $\kappa$. However, one
can be  expressed in terms of elementary functions in some specific
cases. For example, one obtains
\[v(t)=\frac{2k_0\sqrt{u_0}}{\sqrt{b(a-bu_0)}} \left(\arctan\left(\frac{\sqrt{bu_0}}{\sqrt{a-bu_0}}\,e^{\frac{at}{2}}\right)-
\arctan\left(\frac{\sqrt{bu_0}}{\sqrt{a-bu_0}}\right)\right)+v_0\]
in the case $2\alpha=a, \ \kappa=1$.

In \cite{ch-dav-preprint20, ch-dav-2020}, it was demonstrated that
the nonlinear system  (\ref{2-10}) with correctly-specified
parameters and given initial conditions can be successfully
 used for describing the first wave of the COVID-19 pandemic in
several countries (China, Austria, France). In particular, it was
established that the exponent $\kappa$ takes different values in
different countries. For example, $\kappa=1$ (i.e. the case of the
logistic equation in (\ref{2-10}))  leads to  very good
correspondence between the exact solution (\ref{2-11}) and measured
data taken from \cite{meters}. However, the parameter  $\kappa$  was
essentially smaller for many countries in Europe during the first
wave of the COVID-19 pandemic, for example, $\kappa=0.4$ for Austria
and France.

Now we point out that the model (\ref{2-10}) was constructed under
essential simplifications
  of the epidemic process in question. In particular, the model implicitly admits that $u\gg
  v$.
    On the other hand, it is well known that the COVID-19 outbreak
  in several  countries was so severe that the mortality rate  was rather high, i.e.
  the assumption  $u\gg v$  is not true.
   In such cases,  the model (\ref{2-10})   can be generalized  as follows
 \begin{equation}\label{2-13}\ba \frac{du}{dt} =  (u-v)(a-b(u-v)^{\kappa}),\medskip \\
\frac{dv}{dt} = k(t)(u-v), \medskip \\ u(0)=u_0\geq0, \
v(0)=v_0\geq0,
  \ea\end{equation}
  In fact,  the time evolution   of the function $u$  cannot depend on
  the infected  persons who  already died.  Similarly, the number
  of new deaths  cannot depend on the people who  already died.
  Taking into account the equality  $u-v=w$, where $w$ is  the total number of recovered persons,
   the nonlinear  model  (\ref{2-13})
  is reducible to the form
 \begin{equation}\nonumber\ba \frac{dw}{dt} =  w(a-k(t)
 -bw^{\kappa}), \medskip
 \\
 \frac{dv}{dt} = k(t)w, \medskip \\ w(0)=w_0=u_0-v_0\geq0, \quad v(0)=v_0\geq0, \ea\end{equation}
 and its exact solution can be expressed  in the explicit form
 \begin{equation}\label{2-15}\ba
   w(t) =\exp\left(at-\int_0^t \ k(\tau) \,{\rm d}\tau\right)
    \left(w_0^{-\kappa} + b\kappa\,\int_0^t \ \exp\left(\kappa a\tau-\kappa \int_0^\tau \  k(z) \,{\rm d}z \right) \,{\rm d}\tau \right) ^
   {-\frac{1}{\kappa}}, \medskip
  \\
      v(t) = v_0 + \int_0^t \ k(\tau)w(\tau) \,{\rm d}\tau.\ea\end{equation}

Obviously, solution (\ref{2-15}) leads to the  solution
\begin{equation}\nonumber\ba
 u(t) = v_0+ w(t) + \int_0^t \ k(\tau)w(\tau) \,{\rm d}\tau, \medskip \\
  v(t) = v_0 + \int_0^t \ k(\tau)w(\tau) \,{\rm d}\tau.\ea\end{equation}
of the model  (\ref{2-13}).





\section{Classical epidemic models with diffusion in
space}\label{sec-3}

A natural generalization of the classical Kermack--McKendric model
that takes into account the diffusion process reads as
\be\label{3-1}\ba
\frac{\p s}{\p t}=d_s\Delta s-\alpha si, \medskip \\
\frac{\p i}{\p t}=d_i\Delta i+\alpha si-\beta i, \medskip \\
\frac{\p r}{\p t}=d_r\Delta r+\beta i,
 \ea\ee
where $s(t,x)$, $i(t,x)$ and $r(t,x)$ are the susceptible, infective
and recovered population densities at time $t$ in position
$x\in{\mathbb{R}}^n, \ n=1,2,3$, respectively. Here, $\Delta$
denotes the Laplace operator, and $d_s, \ d_i$ and $d_r$ are
diffusion constants. Obviously, having the densities $s(t,x)$,
$i(t,x)$ and $r(t,x)$  and a domain $\Omega \subset {\mathbb{R}}^n$
in which an epidemic is spread,  one can calculate  the numbers of
each sub-population using the formulae \be\label{3-1*}
S(t)=\int_{\Omega}s(t,x)dx, \quad I(t)=\int_{\Omega}i(t,x)dx, \quad
R(t)=\int_{\Omega}r(t,x)dx. \ee

 Because the  three-component  nonlinear
reaction-diffusion system is a complicated object, typically the
systems involving only the first two equations are under study. In
fact, assuming that the conservation law (\ref{2-1**}) is still
valid, the function $ R(t)$ can be easily found using (\ref{3-1*}).
 Pioneering works in which an extensive research
has been conducted to explore the dynamics of travelling waves  of
the two-component epidemic models were published in 1970s--1980s
\cite{noble-1974,kallen-1984,murray-1985}.  In \cite{noble-1974} the
system  (with $d_s=d_i$ and $n=1,2$) \be\label{3-1**}\ba
\frac{\p s}{\p t}=d_s\Delta s-\alpha si, \medskip \\
\frac{\p i}{\p t}=d_i\Delta i+\alpha si-\beta i,
 \ea\ee
was suggested in order  to describe the spread of the well-known
black death pandemic. In particular,  the conditions for the
existence of travelling waves (nowadays the terminology `travelling
fronts' is  used) solutions  are  analyzed, and the numerical
solution in the one-dimensional case $n=1$ are presented.

 In \cite{murray-1985}, the authors consider system
(\ref{3-1**})  with $d_s=0$ and $n=1$ for the spatial spread of
rabies. In \cite{kallen-1984}, a comprehensive analysis of this
model is provided, including a proof of the existence of travelling
waves and  the conditions under which the waves travel at a minimal
speed.

It should be pointed out that  (\ref{3-1**}) is a particular case of
the diffusive Lotka--Voltera system (see, e.g.,
\cite{murray-1989,ch-dav-book}) \be\label{0-4}\ba
\frac{\p u}{\p t} = d_u\Delta u+u(a_1+b_1u+c_1v), \\
\frac{\p v}{\p t} = d_v\Delta v+ v(a_2+b_2u+c_2v), \ea\ee where
 $u=u(t,x)$  and $v=v(t,x)$ are to-be-found
functions, which usually  represent densities, $a_k,\ b_k$ and $c_k
\ (k=1,2)$
 are given parameters.
 Depending on  signs of the above  parameters
 system (\ref{0-4}) describe several types of  interactions
 between species, cells, chemicals, etc. In particular, the prey-predator
 model
 \be\label{0-5}\ba
\frac{\p u}{\p t} = d_u\Delta u+u(a_1-c_1v), \\
\frac{\p v}{\p t} = d_v\Delta v+ v(-a_2+b_2u) \ea\ee
is obtained (here $a_1, \ c_1, \ a_2$ and $b_2$ are positive
constants). Obviously, system (\ref{0-5}) with $a_1=0$ coincides
with (\ref{0-4}) up to notations.

It should be stressed  that construction of  travelling fronts in
explicit forms of the prey-predator system (\ref{0-5}) is a highly
nontrivial problem. To the best of our knowledge, the first examples
of such type  solutions were presented in the recent review
\cite{ch-dav-2022}. Therefore, it is not surprisingly that in the
above-cited works devoted to the epidemic model (\ref{3-1**})
travelling fronts were not found.

The generalization of system (\ref{3-1}) in the case when the
parameters $\alpha$ and $\beta$
 are assumed to be positive
periodic continuous functions in $t$ is considered in paper
\cite{zhang-2020}. In this paper, the existence of periodic
travelling wave solutions  of the form $(s,\ i, \ r)= \left(S(x+\nu
t),I(x+\nu t),R(x+\nu t)\right)$ (here $\nu$ is the wave speed) is
also analyzed. The existence of a travelling wave essentially
depends on the basic reproduction number $R_0$. It is defined by the
formula $R_0=\frac{\alpha}{\beta}S(-\infty)$ in the case of constant
parameters $\alpha$ and $\beta$, and by the formula
$R_0=\frac{\int_0^T\alpha(t)dt}{\int_0^T\beta(t)dt}S(-\infty)$
 in the case of nonconstant parameters.  Note that in
the case  $R_0 \leq 1$ (see, Theorem~3.1~\cite{zhang-2020}), there
are no travelling wave solutions.
This is in agreement with the general statement that a
reaction-diffusion system can describe an  epidemic process and  a
relevant  travelling wave   exists if
 $R_0>1$. This means  that the disease can propagate and
sustain itself in the population. If $R_0 \leq 1$ then the disease
will die out and there will be no travelling wave solutions in the
system.

In \cite{zhi-2023}, the main   focus is  on the problem  how human
behavior can affect the  COVID-19 spread  using a diffusive SEIR
epidemic model.  The model   takes into account contact rate
functions,  describing  different behaviors and interactions among
individuals,  and reads as \be\label{3-14}\ba \frac{\p s}{\p t}=d_s
\frac{\p^2 s}{\p x^2}+\Lambda-\alpha_e(x)\big(a_1 - b_1m_1(e)\big)se
-\alpha_i(x)\big(a_2 - b_2m_2(i)\big)si-\mu_{nd} s, \medskip \\
\frac{\p e}{\p t}=d_e \frac{\p^2 e}{\p x^2}+\alpha_e(x)\big(a_1 -
b_1m_1(e)\big)se+
\alpha_i(x)\big(a_2 - b_2m_2(i)\big)si-\gamma e-\mu_{nd} e, \medskip \\
\frac{\p i}{\p t}=d_i \frac{\p^2 i}{\p x^2}+\gamma e-\beta i-\mu i -\mu_{nd} i, \medskip \\
\frac{\p r}{\p t}=d_r \frac{\p^2 r}{\p x^2}+\beta i - \mu_{nd} r, \\
 \ea\ee
where $\Lambda$ is the influx rate of susceptible individuals;
$\mu_{nd}$ is the natural death rate of the
 human; $\mu$ is the death rate of infected individuals due to
COVID-19; $a_1$ and $a_2$ are the direct contact rates of $e$ and
$i$; $b_1$ and $b_2$ are the associated largest reduced rates due to
human behavior changes of $e$ and $i$; the functions $\alpha_e(x)$
and $\alpha_i(x)$ are the direct transmission contribution rates of
$e$ and $i$, which are probabilities that measure the contribution
of spatial heterogeneity into direct human-to-human transmission,
$\alpha_e(x)$ and $\alpha_i(x)$ and assumed to be nonnegative and
H\"{o}lder continuous functions, e.g., periodic trigonometric
functions; $m_1(e)$ and $m_2(i)$ are saturation functions satisfy
the restrictions
\[a_j\geq b_j>0, \ 0\leq m_j \leq \frac{a_j}{b_j}, \ m_j\in C^1([0,\infty)), \ m_j'\geq0, \
j=1,2.\]

In \cite{zhi-2023}, the basic reproduction number $R_0$ is derived
and a threshold-type result on its global dynamics in terms of $R_0$
is established using the  diffusive SEIR  model (\ref{3-14}). In
order to define $R_0$ for the model in question, the authors analyze
a linear system  near the disease-free steady-state point $E_f$.
Obviously, the nonlinear system (\ref{3-14}) possesses such point of
the form
  $E_f = (s^*, 0, 0, 0)$ with
 $s^*=\frac{\Lambda}{\mu_{nd}}$.
The linearized system (\ref{3-14}) in a vicinity of the equilibrium
point $E_f$ reads as \be\label{3-14-3}\ba
\frac{\p s}{\p t}=d_s \frac{\p^2 s}{\p x^2}-a_1\alpha_e(x)s^*e-a_2\alpha_i(x)s^*i-\mu_{nd}s^*, \medskip \\
\frac{\p e}{\p t}=d_e \frac{\p^2 e}{\p x^2}+a_1\alpha_e(x)s^*e+a_2\alpha_i(x)s^*i-(\gamma+\mu_{nd})e, \medskip \\
\frac{\p i}{\p t}=d_i \frac{\p^2 i}{\p x^2}+\gamma e-(\beta+\mu+\mu_{nd})i, \medskip \\
\frac{\p r}{\p t}=d_r \frac{\p^2 r}{\p x^2}+\beta i - \mu_{nd} r.\\
 \ea\ee
Since the equations for $e$ and $i$ do not involve  $s$ and $r$, the
authors consider the following subsystem: \be\label{3-14-4}\ba
\frac{\p e}{\p t}=d_e \frac{\p^2 e}{\p x^2}+a_1\alpha_e(x)s^*e+a_2\alpha_i(x)s^*i-(\gamma+\mu_{nd})e, \medskip \\
\frac{\p i}{\p t}=d_i \frac{\p^2 i}{\p x^2}+\gamma e-(\beta+\mu+\mu_{nd})i,  \\
 \ea\ee
and derive the formula for the basic reproduction number
\be\label{3-14-5}R_0=sup\{|\lambda|:\,\lambda\in\sigma(L)\}.\ee In
(\ref{3-14-5}), $\sigma(L)$ is the spectral set of operators $L$
defined as \be\label{3-14-6}L(\phi)(x)=F(x)\int_0^\infty
T(t)\phi(x)dt.\ee Here $\phi=(\phi_2,\phi_3)\in
C\left(\bar{\Omega},\mathbb{R}^2\right)$ represents the initial
distribution of the  densities $e$ and $i$, while the function
$F(x)$ and the operator $T(t)$ in (\ref{3-14-6}) are  defined by
using  the  linear system (\ref{3-14-4}).

  Furthermore,   it is proved that  the disease-free state $E_f$ is stable if
$R_0\leq1$, meaning the infection is not sustained in the
population. However, if $R_0>1$, then a positive stationary solution
exists, indicating that the epidemic can persist and spread
throughout the population. To  investigate further the impact of
human behavior, the authors conduct numerical simulations based on
their analytical findings. These simulations demonstrate that
changes in human behavior can have a positive effect in reducing the
infection level by decreasing the number of infected persons.
 Overall, the study emphasizes the importance
of considering human behavior in modelling the spread of COVID-19
and highlights the potential effectiveness of behavior changes in
mitigating the infection levels and reducing the overall impact of
the pandemic.

In \cite{ahmed-2021}, a reaction-diffusion model \be\label{3-30}\ba
\frac{\partial s}{\partial t} = d_s \frac{\partial^2 s}{\partial
x^2} + \mu_{nd}
 - \alpha_{i}\frac{ s\,i}{1+q i^2} - \mu_{nd} s, \medskip\\
\frac{\partial e}{\partial t} = d_{e} \frac{\partial^2 e}{\partial
x^2} +
\alpha_{i}\frac{ s\,i}{1+q i^2} - (\gamma + \mu_{e} + \epsilon + \mu_{nd})e, \medskip\\
\frac{\partial i}{\partial t} = d_{i} \frac{\partial^2 i}{\partial
x^2} +
 \gamma e - (\sigma + \mu_{i} + \mu_{nd})i, \medskip\\
\frac{\partial r}{\partial t} = d_r \frac{\partial^2 r}{\partial
x^2} + \sigma i + \epsilon e - \mu_{nd} r
 \ea\ee
for the spread of COVID-19 is investigated. The model is a spatial
extension of the SEIR model with nonlinear incidence rates by taking
into account the effects of random movements of individuals from
different compartments (subpopulations). The diffusive SEIR model
(\ref{3-30}) can be considered as a very particular case of model
(\ref{3-14}). Indeed, the authors use the specific function $\frac{
1}{1+q i^2}$ instead of a general function $m_2(i)$ (see
(\ref{3-14})). However, it can be noted  that system (\ref{3-30})
contains new term $\epsilon\,e$ that takes into account   the
COVID-19 immunity of exposed individuals.

In system (\ref{3-30}),  $q$ is the bilinear incidence rate;
$\epsilon$ is  the immunity rate of exposed individuals; $\sigma$ is
the rate of vaccination, quarantine or treatment; $\mu_{e}$ is the
mortality rate of exposed individuals due to virus; $\mu_{i}$ is the
death rate of infected individuals due to virus. All other
parameters have the same interpretations as above.

The results presented in \cite{ahmed-2021} can be briefly summarized
as follows: the equilibrium points and their stability are
investigated; stability regions and influence of key parameters are
explored; a structure-preserving finite difference method for
simulating the  model in question is developed; consistency and
stability analysis  is provided,  positivity of solutions is
discussed.

In  \cite{mammeri-2020}, a model for the spread of the COVID-19
epidemic is constructed based on the diffusive SEIR model  by
incorporating so-called asymptomatic infections. This model consists
of three reaction-diffusion equations  and two ODEs (see
(1)~\cite{mammeri-2020}) \be\label{3-15}\ba
\frac{\p s}{\p t}=d_s(t) \triangle s-f(t)\left(\alpha_ee+\alpha_{i_s}i_s+\alpha_{i_a}i_a\right)\frac{s}{N}, \medskip \\
\frac{\p e}{\p t}=d_s(t) \triangle e+f(t)\left(\alpha_ee+\alpha_{i_s}i_s+\alpha_{i_a}i_a\right)\frac{s}{N}-\gamma e, \medskip \\
\frac{\p i_a}{\p t}=d_s(t) \triangle i_a+(1-p)\gamma e-\beta i_a, \medskip \\
\frac{\p i_s}{\p t}=p\gamma e-\beta i_s-\mu i_s, \medskip \\
\frac{\p r}{\p t}=\beta(i_a+i_s).
 \ea\ee
In system (\ref{3-15}), the functions $i_a$ and $i_s$ are the
densities of asymptomatic and symptomatic infected individuals,
respectively; the diffusion $d_s(t)$ has one of the forms presented
in formulae (3) \cite{mammeri-2020}; the function $f(t)$ represents
the average number of contacts (see (2) in \cite{mammeri-2020}); the
parameter $p$ is the probability of being confirmed, while $(1-p)$
is the probability of being unreported; other parameters have the
same interpretations as above.

Using official data, spatial modelling of the density distribution
of symptomatic infected individuals in France is performed during
the first wave of the pandemic (from January 24th to June 16th,
2020). The computational results show good agreement with the
official data. It is demonstrated that the total number of cases
would be significantly higher without intervention of the government
(i.e. without implementing a lock-down).


In  \cite{zhu-2019,tu-2023}, reaction-diffusion models are proposed
to investigate the impact of vaccination and isolation strategies on
the progression of the epidemic. In \cite{zhu-2019}, the global
asymptotic stability and the persistence of the epidemic are proven
using a reaction-diffusion model for the HBV epidemic. Results of
numerical simulations are presented as well.

The study  \cite{tu-2023} begins by exploring the fundamental
dynamic properties of the diffusive epidemic system. Subsequently,
the asymptotic distributions of the endemic equilibrium under
different conditions are analyzed. Overall, this research
contributes to the ongoing efforts in epidemic prevention and
control by providing insights into the dynamics of COVID-19 and
suggesting optimal vaccination and isolation strategies. The
relevant   model of the COVID-19 epidemic was constructed taking
into account the vaccination of patients. The model is based on the
system of four reaction-diffusion equations for a heterogeneous
medium and has the form \be\label{3-21}\ba
\frac{\p s}{\p t}=d_s \Delta s+f_1 (x)-\alpha_1 (x)\frac{ k }{1+a_1 k i}si-[\sigma(x)+\mu_{nd}(x)]s + \omega(x)r, \medskip \\
\frac{\p v}{\p t}=d_v \Delta v+f_2 (x)-\alpha_2 (x)\frac{ k }{1+a_2 k i}vi+\sigma(x)s-\mu_{nd}(x)v , \medskip \\
\frac{\p i}{\p t}=d_i \Delta i+\alpha_1 (x)\frac{ k }{1+a_1 k i}si+\alpha_2 (x)\frac{ k }{1+a_2 k i}vi-[\beta(x)+\mu(x)+\mu_{nd}(x)]i , \medskip \\
\frac{\p r}{\p t}=d_r \Delta r+\beta(x)i-[\omega(x)+\mu_{nd}(x)]r , \medskip \\
 \ea\ee
where the densities $s(t,x)$, $v(t,x)$, $i(t,x)$ and $r(t,x)$ stand
for unvaccinated susceptible individuals, vaccinated susceptible
individuals, infected individuals, and recovered individuals,
respectively; $k=1-m, \ m\in [0, 1)$ (here $m$ means the fraction of
infected subpopulation). It is assumed that all parameters and
functions in (\ref{3-21}) are nonnegative. The parameters $f_1(x)$
and $f_2(x)$  are the inputs to $s$ and $v$
 respectively;
 $a_1$ and $a_2$ are so-called half-saturation
parameters; $\sigma(x)$ denotes the vaccination rate of $s$;
$\omega(x)$ indicates the immune loss rate of $r$; other parameters
have the same interpretations as above.


To construct numerical solution of the reaction-diffusion model
(\ref{3-21}), the finite difference method in time and space is
used. Plots of the numerical solutions are presented in the 1D space
approximation.  The authors pay special attention to the cases when
one or more diffusivities are small. For numerical simulations, the
parameters and functions in (\ref{3-21}) were specified using
available experimental data from a wide range of references. Based
on the obtained numerical results,  the sensitivity index of the
relevant parameters for the basic reproduction number $R_0$ is
obtained. It is shown that the functions $f_1(x)$, $\beta_1(x)$,
$\alpha_2(x)$ and  $\beta(x)$ are highly sensitive parameters, while
the functions $\mu_{nd}(x), \ \omega(x)$ and $\mu(x)$ are
insensitive parameters. The sensitivity of the parameters provides
technical guidance for COVID-19 prevention and control. Controlling
the high sensitivity parameter, one can better reduce the basic
reproduction number~$R_0$.

It should be pointed out that all the models presented above involve
constant (or time-dependent) diffusivities.
 Recently, several papers were published \cite{viguerie-2020,grave-2021,viguerie-2021,grave-2022} by an international group of researchers in which a new
  SEIRD type model with nonlinear  diffusion terms was   introduced   and examined by analytical and numerical methods.
 In the simplest case the model has the form \cite{viguerie-2020,grave-2021}
  \be\label{3-3**}\ba
\frac{\p s}{\p t}=\nabla\cdot\left(d_sN\nabla e\right)-\frac{\alpha}{N} si, \medskip \\
\frac{\p e}{\p t}=\nabla\cdot\left(d_eN\nabla e\right)+\frac{\alpha}{N} si-\gamma e, \medskip \\
\frac{\p i}{\p t}=\nabla\cdot\left(d_iN\nabla i\right)+\gamma e-\beta i-\mu i, \medskip \\
\frac{\p r}{\p t}=\nabla\cdot\left(d_rN\nabla r\right)+\beta i, \medskip \\
\frac{\p d}{\p t}=\mu i.
 \ea\ee
Hereinafter $N(t,x)$ denotes the sum of the living population
densities, i.e., $N=s+e+i+r$; the diffusion parameters $d_s, \ d_e,
\ d_i$ and $d_r$ may depend on time and space. Here all diffusivity
coefficients are proportional to the total population and can be
locally adjusted to incorporate geographical or human-related
inhomogeneities (see Section 3 \cite{keller-2013} for details).

In \cite {grave-2021}, a further generalization of the above  model
is suggested:
\be\label{3-3}\ba
\frac{\p s}{\p t}=\nabla\cdot\left(d_sN\nabla s\right)-\lf(1-\frac{A}{N}\rg)\alpha_isi-\lf(1-\frac{A}{N}\rg)\alpha_ese, \medskip \\
\frac{\p e}{\p t}=\nabla\cdot\left(d_eN\nabla e\right)+\lf(1-\frac{A}{N}\rg)\alpha_isi+\lf(1-\frac{A}{N}\rg)\alpha_ese-\gamma e-\beta_ee, \medskip \\
\frac{\p i}{\p t}=\nabla\cdot\left(d_iN\nabla i\right)+\gamma e-\beta_i i-\mu i, \medskip \\
\frac{\p r}{\p t}=\nabla\cdot\left(d_rN\nabla r\right)+\beta_i i+\beta_e e,\medskip \\
\frac{\p d}{\p t}=\mu i,
 \ea\ee
where a new parameter $A$ is used to describe the Allee effect
(depensation), which serves for describing  a tendency of outbreaks
to cluster towards small population centers. The Allee effect is a
phenomenon in population dynamics that describes a decrease in the
per capita growth rate of a population as it approaches lower
population densities. In the context of disease outbreaks, the Allee
effect is often used to model the tendency for outbreaks to occur
more frequently and cluster in smaller population centers or areas
with lower population densities.

The diffusive SEIRD system (\ref{3-3}) is solved  using finite
element methods \cite{grave-2021}. To verify the model accuracy, the
obtained results are cross-referenced with the research outcomes of
other authors.

In  \cite{viguerie-2020,viguerie-2021}, a modification of the
diffusive SEIRD system (\ref{3-3}) by incorporating the general
(nondisease) mortality rates is studied.
 In the modified model,  the linear terms $-\mu_{nd} s, \ -\mu_{nd} e, \
-\mu_{nd} i$ and $-\mu_{nd} r$ are present on the
 right-hand side of the first four equations of system
(\ref{3-3}), respectively.

In  \cite{viguerie-2021},  the diffusive SEIRD model is used to
describe the spatiotemporal spread of the COVID-19 pandemic and aims
to capture dynamics based on human behaviors and geographical
features. To validate the model, numerical results are compared with
measurement data from the Lombardy region in Italy, which was
severely affected by the crisis between February and April 2020. The
obtained results demonstrate qualitative agreement between the
modeled spatiotemporal spread of COVID-19 in Lombardy and
epidemiological data.
 It is concluded that the numerical  results
can be used to inform healthcare authorities in developing effective
measures to mitigate the pandemic and anticipate the geographical
distribution of critical medical resources.

In  \cite{viguerie-2020}, the authors analized an ODE version of the
diffusive SEIRD model to derive a basic reproduction number $R_0$.
Additionally, the authors explored the role of diffusion and $R_0$
in shaping the behavior of the diffusive SEIRD model. Through
numerical simulations, they investigated how the interplay between
these factors affects the dynamics of the epidemic.

In order to show applicability of the model, the role of diffusion
is  demonstrated  in the case  of the Lombardy region (Italy).
It is shown that the mathematical model (\ref{3-3**}) reproduces the
COVID-19 epidemic spread of the  in Lombardy, beginning on February
27, 2020.  In addition to earlier  simulations presented in
\cite{viguerie-2021}, two additional cases are investigated in
\cite{viguerie-2020}. In the first case, the values of $d_s$, $d_e$,
$d_i$, and $d_r$ are doubled, while in the second case, they are
halved. Furthermore, a scenario is  examined in which $d_r$, $d_e$,
and $d_i$ are doubled, but $d_s$ is halved. This choice closely
resembles the parameter configuration used in the 2D simulations.
The primary objective is to prevent the potential occurrence of
nonphysical diffusion within the susceptible population, which may
result in a general decrease of the  population density. It is also
observed that a wider geographic range of affected areas is produced
by larger diffusion what is in agreement with physical meaning of
diffusion. This effect is particularly evident in the southeastern
clusters (see Fig.~11~\cite{viguerie-2020}). In the case of double
diffusion, a homogeneous and continuous region of infection is
generated. In contrast,  more localized dynamics are observed in the
case of half diffusion, resulting in a clear separation into
distinct regions. The  case, in which $d_r$, $d_e$, and $d_i$ are
doubled and  $d_s$ is halved, produces intermediate results between
the double-diffusion and half-diffusion cases.

\section{Other  epidemic models taking into account
spatial heterogeneity }\label{sec-4}

There are several models for simulating  epidemic spreads in time
and space that cannot be treated as direct generalisations of the
classical models presented in Section~\ref{sec-2} by adding
diffusion terms. Some of them are presented in this section. It
should be stressed that almost all of them were developed after the
COVID-19 outbreak.

We start from the simplest  model of such type  that was introduced
in \cite{cheng-2022}.  The model reads as  \be\label{4-1}\ba
\frac{\p s}{dt}=-\alpha si, \medskip \\
\frac{\p i}{dt}+\nabla\cdot\left(iV\rg)=\alpha si-(\beta+\mu)i, \medskip \\
\frac{\p r}{dt}=\beta i,
 \ea\ee
where the functions $s, \ i, \ r$ and the parameters $\alpha,\
\beta$ and $\mu$ have the  same interpretation as in the previous
section.
 Spatial movement for the
infected population relative to the medium is investigated by
introducing the function
 $V(t,x)$
 satisfying the Euler equation \be\label{4-2} i\frac{\p V}{\p t}+
iV\cdot\nabla V=-\nabla p,\ee  where $p$ is the pressure, which is
an internal force of the fluid and assumed to be a known smooth
function. It should be noted that the model (\ref{4-1})--(\ref{4-2})
consists of the first-order equations in contrast
 to the models presented in Section~\ref{sec-3}.
 In particular, it means that much simpler equations are obtained for search for
  travelling waves. Moreover, it can be noted that an autonomous  system for finding  the functions $i$ and $V$ consisting of the first-order PDEs
(\ref{4-2}) and
\[ \nabla\cdot\left(iV\rg)=-\mu i          \]
is obtained providing the conservation law (\ref{2-1**}) is
preserved for the densities
 $s, \ i, \ r$.

 In model  (\ref{4-1})--(\ref{4-2}), the motion of the infected population, which
characterizes the spatial spread of the epidemic, is described as an
inviscid fluid. It is important to note that there is no requirement
to assume the susceptible population remains stationary in this
modelling framework. Instead of explicitly focusing on the movement
of susceptible individuals, the susceptible population is treated as
a medium, and the motion of the infected fluid is analyzed in
relation to this medium. While individuals may exhibit diverse and
random movements at the individual level, this model is based on the
underlying assumption that the spread of an epidemic can be
approximated by an inviscid flow at the macroscopic level.
 As an illustrative demonstration of model applicability,
 the spread of the COVID-19 epidemic  within Wuhan, China 2020, was
 used.

To find the numerical solution of model (\ref{4-1})--(\ref{4-2}),
the finite difference method was applied. The difference scheme
obtained was tested using a known analytical solution. Having done
this, the numerical simulation were performed taking into account
the number of residents and the area of Wuhan. The authors modeled
the spread of the epidemic over a square area equal to the area of
the city  assuming that the source of infection is located in the
central part of the city. The results obtained by numerical
simulations are in good agreement with the official data of the
COVID-19 disease in Wuhan.

In paper \cite{Zhuang-2021}, a spatio-temporal model of the spread
of the COVID-19 epidemic with moving boundaries was constructed. The
governing equations are direct generalization of those presented in
model (\ref{4-1})--(\ref{4-2}). As a result, a system of four
equations of the SEIR model with convective terms was obtained,
where the speed of movement of individuals is taken into account.
The model reads as \be\nonumber\ba
\frac{\p s}{dt}+\nabla\cdot\left(sV\rg)=\Gamma-\alpha_e se-\alpha_i si, \medskip \\
\frac{\p e}{dt}+\nabla\cdot\left(eV\rg)=\alpha_e se+\alpha_i si-\gamma e, \medskip \\
\frac{\p i}{dt}+\nabla\cdot\left(iV\rg)=\gamma e-(\mu+\beta)i, \medskip \\
\frac{\p r}{dt}+\nabla\cdot\left(rV\rg)=\beta i,
 \ea\ee where the function $V$ is the speed that
characterizes the epidemic flow; the function $\Gamma$  represents
the rate of the density change for the susceptible individuals due
to the expansion of the epidemic domain; all other parameters have
the same interpretations as above. The functions $V$ and $\Gamma$
generally depends on both the time and space, while all other
parameters are assumed to be constants.

The authors assume that there are no infected individuals outside a
2D space domain with the  boundary $Z$. Because one needs to
introduce some assumptions about the moving boundary $Z$, the
authors studied radially-symmetric case. In this case the model
essentially simplifies, in particular, the boundary $Z$ is nothing
else but a circle with the time depended radius $L$. Thus, using
rather a standard approach, the equations for $L$ and the speed $V$
were derived (see (2.6) and (2.7) in \cite{Zhuang-2021}). The model
is verified on official data on the epidemic in Wuhan for the time
frame: from January 23 to February 16--18, 2020. Based on the
results of calculations, the radial distribution of speed and
infected individuals are found.

A model involving cross-diffusion was studied in
\cite{ch-dav-2022-cd}.
 The model is a generalization of the ODE  model (see  (\ref{2-10}) above)) developed in  \cite{ch-dav-2020}  and reads as
\begin{equation}\label{3-4}\begin{array}{l} \medskip  \frac{\p u}{\p t} = d_1 \Delta u+u(a- bu^\kappa),  \\
\frac{\p v}{\p t} = d_2\Delta u +k(t)u.\end{array}\end{equation}
In (\ref{3-4}), the function $u(t,x,y)$ describes the density (rate)
of the infected persons (the number of the  COVID-19 cases) in a
vicinity of the point $(x,y)$,  while $v(t,x,y)$ means the density
of the deaths from COVID-19. The diffusivity coefficients $d_1$ and
$d_2$ describe the random movement  of the infected persons, which
lead to increasing the pandemic spread. Each  coefficient in the
reactions terms, $a, \ b, \ \kappa$  and $k(t)$, has the clear
meaning described above for the ODE  model  (\ref{2-10}).

 It turns out that a wide range of exact solutions (including travelling fronts type those) of the nonlinear system  (\ref{3-4}) can be constructed
 using the Lie symmetry analysis.  We have also demonstrated
 that the exact solutions  obtained  are useful for  describing
 the spread of the COVID-19 pandemic in 1D approximation.

Let us present some details. Taking into account the assumptions
that the distribution of the infected persons is one-dimensional in
space  (i.e., the diffusion w.r.t. the axis $y$ is very small) and
that $d_1 \gg d_2$ (i.e., the space  diffusion of the infected
persons leads mostly to increasing the total  number of the COVID-19
cases and not so much to new deaths), system (\ref{3-4}) with
$k(t)=k_0 e^{-\alpha t}$ (hereafter $k_0>0, \  \alpha>0$) takes the
form
\begin{equation}\label{3-5}\begin{array}{l} \medskip  \frac{\p u}{\p t} = \frac{\p^2 u}{\p x^2}+u\left(1- u^\kappa\right),  \\
\frac{\p v}{\p t} = \frac{k_0}{a}\exp\left(-\frac{\alpha
t}{a}\right)u.
\end{array}\end{equation}
System (\ref{3-5}) admits the Lie symmetry operator (see case
\emph{4)} in Theorem~2.3 \cite{ch-dav-2022-cd}) \[ X = c\partial_x +
\partial_t-\frac{\alpha }{a}\,v\partial_v, \ c \in \mathbb{R}, \]
that allowed us to construct its exact solution in the following
form
\begin{equation}\ba\label{3-6}
u(t,x) =\left(1+ A\exp
\left(\frac{\kappa}{\sqrt{2(\kappa+2)}}\,\omega\right)\right)^{-2/\kappa},
\  \omega=\pm\, x-\frac{\kappa+4}{\sqrt{2(\kappa+2)}}\,t,  \  A>0,
\medskip
\\
v(t,x) =\frac{k_0}{a}\int \exp\left(-\frac{\alpha
t}{a}\right)\left(1+ A\exp
\left(\frac{\kappa}{\sqrt{2(\kappa+2)}}\,\omega\right)\right)^{-2/\kappa}dt
+ g(x), \ea\end{equation} where $g(x)$  is an arbitrary smooth
function.

\begin{remark}
The assumption  $d_1 \gg d_2$ is not essential and one can construct
exact solutions of the form  (\ref{3-6}) for  (\ref{3-4}) in
one-dimensional approximation as well. So, setting $d_1=1, \
d_2=D>0$, the exact solution
\begin{equation}\nonumber\ba
u(t,x) =\left(1+F\right)^{-2/\kappa},
\medskip \\
v(t,x) =\frac{k_0}{a}\int \exp\left(-\frac{\alpha
t}{a}\right)\left(1+F\right)^{-2/\kappa}dt - \frac{D}{2+\kappa}\int
F\left(\kappa-2 F\right)\left(1+ F\right)^{-2-2/\kappa}dt+g(x),
\ea\end{equation} can be  found. Here $F= A\exp
\left(\frac{\kappa}{\sqrt{2(\kappa+2)}}\,\omega\right), \
\omega=\pm\, x-\frac{\kappa+4}{\sqrt{2(\kappa+2)}}\,t,  \  A>0$.
\end{remark}

The integral in (\ref{3-6}) can be expressed in the terms of
elementary functions in some particular cases.  Taking
$\alpha=\frac{5}{6}\,a$ and $\kappa=1$ , for example, one can obtain
the exact solution
\begin{equation}\label{3-7}\begin{array}{l} \medskip
u(t,x) =\left(1+ A\exp \left(\frac{1}{\sqrt{6}}\,\omega\right)\right)^{-2}, \  \omega=x-\frac{5}{\sqrt{6}}\,t,  \\
v(t,x) =g(x) -\frac{6k_0}{5a}\exp\left(-\frac{x}{\sqrt 6}\right)
\left(A+ \exp \left(-\frac{1}{\sqrt{6}}\,\omega\right)\right)^{-1},
\end{array}\end{equation}
of system (\ref{3-5}).

Note that the functions $u$ and $v$ in (\ref{3-7}) should be
nonnegative for any $t>0$ and $x \in \mathbb{I}$ (here $\mathbb{I}
\subset \mathbb{R}$) because they represent the densities.
Obviously, the functions $u$ is always positive.   It is easily seen
that each function $g(x)$ satisfying the inequality
\[  g(x) \geq  \frac{6k_0}{5a}\exp\left(-\frac{x}{\sqrt 6}\right)
\left(A+ \exp \left(-\frac{x}{\sqrt{6}}\right)\right)^{-1} \]
guarantees also  nonnegativity of $v$. In particular, one may take
the function
\begin{equation}\nonumber
 g(x) = \frac{6k_0}{5a}\exp\left(-\frac{x}{\sqrt 6}\right)
\left(A+ \exp \left(-\frac{x}{\sqrt{6}}\right)\right)^{-1},
\end{equation}
which guarantees that the zero density of the deaths in the initial
time $t=0$, i.e., $v(0,x)=0$.

Examining the space interval $\mathbb{I}=[x_1,x_2], \ x_1<x_2$, one
can calculate the total number of the COVID-19 cases and deaths on
this interval as follows
\begin{equation}\label{3-13}\begin{array}{l} \medskip
U(t)=\int_{x_1}^{x_2} u(t,x)dx, \\
V(t)=\int_{x_1}^{x_2} v(t,x)dx.
\end{array}\end{equation}
So, substituting solution (\ref{3-7}) into (\ref{3-13}), we arrive
at the formulae
\begin{equation}\nonumber\begin{array}{l} \medskip
U(t)=(x_2-x_1) - \sqrt{6}\Big[\Big(1+ A\exp \Big(\frac{x_1-\frac{5}{\sqrt{6}}\,t}{\sqrt{6}}\Big)\Big)^{-1}-\Big(1+ A\exp \Big(\frac{x_2-\frac{5}{\sqrt{6}}\,t}{\sqrt{6}}\Big)\Big)^{-1}\\
\hskip2cm +\ln\Big(1+ A\exp \Big(\frac{x_2-\frac{5}{\sqrt{6}}\,t}{\sqrt{6}}\Big)\Big)- \ln\Big(1+ A\exp \Big(\frac{x_1-\frac{5}{\sqrt{6}}\,t}{\sqrt{6}}\Big)\Big)\Big],\medskip \\
V(t)=  \int_{x_1}^{x_2} g(x)dx  -
\frac{6\sqrt{6}\,k_0}{5a}e^{-\frac{5}{6}\,t}\Big[ \ln\Big(1+ A\exp
\Big(\frac{\frac{5}{\sqrt{6}}\,t-x_1}{\sqrt{6}}\Big)\Big)
-\ln\Big(1+ A\exp
\Big(\frac{\frac{5}{\sqrt{6}}\,t-x_2}{\sqrt{6}}\Big)\Big].
\end{array}\end{equation}
Obviously the functions $U(t)$  and $V(t)$  are increasing and
bounded, because
\[ (U,V) \to \Big((x_2-x_1),  \int_{x_1}^{x_2} g(x)dx \Big) \  as \  t \to +\infty. \]
Moreover, taking
the  appropriate function $g(x)$, we  can guarantee that
\[ U(0)=U_0\geq 0,  \  V(0)=V_0\geq 0.\]
Thus, one may claim that the exact solution  (\ref{3-7}) possesses
all  necessary properties for the description of the  distribution
of the COVID-19 cases (total number of infected population as well)
and the deaths from this virus in time-space.

Interestingly, the spread of the COVID-19 cases in space has the
form of a travelling wave, and this qualitatively coincides  with
the real situation in many countries during the first wave of the
COVID-19 pandemic. In Ukraine, for example, the pandemic started in
the western part of the country and then spread to the central and
eastern parts of Ukraine (the major exception was only the capital
Kyiv, in which the total number of COVID-19 cases was high from the
very beginning).

It should be stressed that cross-diffusion  is a distinguished
peculiarity in some well-known  mathematical models arising in
biology, ecology and medicine \cite{kel-seg-71, sh-ka-te} that were
extensively studied by different mathematical techniques (see, e.g.
\cite{ch-da-ki-23} and references therein), during the last decade.
So, we believe that relevant terms with cross-diffusion should
naturally arise in epidemic models. To the best of our knowledge,
the first epidemic model with cross-diffusion was briefly  presented
in \cite{bailey-1975} (see Section 9.2 therein) that reads as
 \be\label{4-1ad}\ba
\frac{\p s}{\p t}=d_s\Delta s-d_{is}s\Delta i-\alpha si+b, \medskip \\
\frac{\p i}{\p t}=d_i\Delta i+d_{is}s\Delta i+\alpha si-\beta i,
 \ea\ee
 where $d_{is}>0$ is cross-diffusivity and the term $d_{is}s\Delta i$  reflects infection caused by isotropic movement in space (in contrast to the term
  $\alpha si$ reflecting a local infection). Here also introduced
   a birth parameter $b$, however, we believe that this parameter should be a function of $s$, the simplest possibility is $b=b_1s, \ b_1>0$.  Moreover, a similar term, say $b=b_2i, \ b_2>0$,  should be added in the second equation because the infected population can also produce new members. So, the above model with cross-diffusion  (\ref{4-1ad}) can be generalized as follows
   \be\label{4-2ad}\ba
\frac{\p s}{\p t}=d_s\Delta s-d_{is}s\Delta i-\alpha si+b_1s, \medskip \\
\frac{\p i}{\p t}=d_i\Delta i+d_{is}s\Delta i+\alpha si +(b_2-\beta)
i.
 \ea\ee
It can be noted that the nonlinear model (\ref{4-2ad}) can be
essentially simplified in the special case $d_s=d_i $ (a plausible
assumption) and $b_1=b_2-\beta$ when one of the equations in
(\ref{4-2ad}) can be replaced by the linear reaction-diffusion
equation
 \be\nonumber
\frac{\p z}{\p t}=d_s\Delta z+b_1z, \quad  z=s+i. \ee

\section{Age-structured epidemic models }\label{sec-5}

In  this section, we  present some information about so-called
age-structured epidemic models.  The simplest representative of such
type models reads as
 \be\label{4-4}\ba
 \frac{\p s}{\p t}+\frac{\p s}{\p a} =-\lambda(a,t) s-\mu_{nd}(a)s, \medskip \\
 \frac{\p i}{\p t}+\frac{\p i}{\p a} =\lambda(a,t) s-\beta(a)i-\mu_{nd}(a) i, \medskip \\
 \frac{\p r}{\p t}+\frac{\p r}{\p a} =\beta(a) i-\mu_{nd}(a)r,
 \ea\ee where $s(a,t), \ i(a,t)$ and $r(a,t)$ are the age-specific density of susceptible, infective
and recovered individuals of age $a$ at time $t$, respectively; the
function $\lambda(a,t)$ is the age-specific force of infection (the
probability for a susceptible of age $a$ to become infective in a
unit time interval);
 $\mu_{nd}(a)$ and $\beta(a)$ are the age-specific death and recovery
 rates, respectively.

 Depending on the form of the function
$\lambda(a,t)$ system (\ref{4-4}) can be a system of the
two-dimensional first-order PDEs or a system of integro-differential
equations. System (\ref{4-4}) with
\[\lambda(a,t)=k(a)\,i(a,t) \quad \texttt{and} \quad
\lambda(a,t)=k(a)\int_0^{\infty} i(a,t)da\] (here $k(a)$ is
nonnegative bounded continuous function on $[0,\infty)$) was
suggested in  \cite{greenhalgh-1987,busenberg-1988}. In particular,
 an endemic threshold criteria is derived and the stability of
steady-state solutions of system (\ref{4-4}) is determined therein.

System (\ref{4-4}) with a more general form of the  function
$\lambda(a,t)$ on the the interval $[0, a_{max}]$ (one is more
realistic) is studied in \cite{inaba-1990}. Setting
\[\lambda(a,t)=\int_0^{a_{max}}k(a,\sigma)i(\sigma,t)d\sigma\] (here $k(a,\sigma)$ is the age-dependent transmission
coefficient, i.e., the probability that a susceptible person of age
$a$ meets an infectious person of age $\sigma$ and becomes infected,
per unit of time,  $a_{max}$ is a maximal age)
 conditions that
guarantee the existence and uniqueness of nontrivial steady-states
of the age-structured model are  derived, the local and global
stabilities of the steady-states are  examined.

The age-structured SIR model (\ref{4-4}) with
$\lambda(t)=\int_0^{\infty}k(a)i(a,t)da$ is studied in
\cite{hadeler-2017} (see Section 6.4 therein). In particular, the
basic reproduction number $R_0$ is obtained (see formula 6.72
therein). A  limiting case of  the age-structured SIR model
(\ref{4-4}) is  called   the age-structured SIS model and that was
studied in detail.  The model reads as \be\label{4-7}\ba
 \frac{\p s}{\p t}+\frac{\p s}{\p a} =-\lambda\frac{\alpha(a)}{N}\, s+\beta(a) i-\mu_{nd}(a)s, \medskip \\
 \frac{\p i}{\p t}+\frac{\p i}{\p a} =\lambda\frac{\alpha(a)}{N}\, s-\beta(a)i-\mu_{nd}(a)
 i, \medskip\\
 s(0,t)=\int_0^{\infty}b(a)\left(s+i\right)da, \
 i(0,t)=0, \medskip \\
\lambda=\int_0^{\infty}k(a)i(a,t)da, \
N=\int_0^{\infty}\left(s+i\right)da.
 \ea\ee
 Here  $N(t)$ is the total population, $b(a)$ is the birth
rate. The exact solution of the age-structured SIS model (\ref{4-7})
is constructed in the  form \cite{hadeler-2017} \be\label{4-8}
s(a,t)=S(a)e^{\kappa t}, \ i(a,t)=I(a)e^{\kappa t}, \ee where
$\kappa$ is to-be-determined constant. Using ansatz (\ref{4-8}),
 the age-structured
SIS model (\ref{4-7}) is reduced to the following form
\be\label{4-9}\ba
 \frac{d S}{d a} =-\lambda^*\frac{\alpha(a)}{N^*}\, S+\beta(a) I-\mu_{nd}(a)S-\kappa S, \medskip \\
  \frac{d I}{d a} =\lambda^*\frac{\alpha(a)}{N^*}\, S-\beta(a)
  I-\mu_{nd}(a)I -\kappa I, \medskip\\
 S(0)=\int_0^{\infty}b(a)\left(S+I\right)da, \
 I(0)=0, \medskip\\
 \lambda^*=\int_0^{\infty}k(a)I(a)da, \ N^*=\int_0^{\infty}\left(S+I\right)da.
 \ea\ee
Formulae  (\ref{4-9}) form a boundary-value problem with the
governing equations in the form of ordinary integral-differential
equations. Using the governing  equations  from  (\ref{4-9}), one
arrives at the relation \be\label{4-10}S(a)+I(a)=C\exp\left(-\kappa
a -\int_0^a\mu_{nd}(\sigma)d\sigma\right).\ee  Assuming
$S(0)+I(0)=1$ (without losing a generality), one may set $C=1$.
Thus, using (\ref{4-10}) and the boundary  conditions from
(\ref{4-9}), the parameter  $\kappa$ is defined by the transcendent
equation \be\label{4-13}\int_0^{\infty}b(a)\exp\left(-\kappa a
-\int_0^a\mu_{nd}(\sigma)d\sigma\right)da=1.\ee Integrating the
second equation from (\ref{4-9}) and taking into account the above
formulae, the exact solution of the boundary-value problem
(\ref{4-9})
 was found \cite{hadeler-2017}
\be\label{4-11}\ba S(a)=\exp\left(-\kappa a
-\int_0^a\mu_{nd}(\sigma)d\sigma\right)-I(a),  \medskip \\
 I(a)=\lambda^*\exp\left(-\kappa a
-\int_0^a\mu_{nd}(\sigma)d\sigma\right)\int_0^a\frac{\alpha(\sigma)}{N^*}\exp\left(-\int_\sigma^a\beta(\varsigma)d\varsigma
-\lambda^*\int_\sigma^a\frac{\alpha(\varsigma)}{N^*}d\varsigma\right)d\sigma,
 \ea\ee where $\lambda^*$ satisfies the equation
\be\label{4-12}\int_0^{\infty}k(a)\exp\left(-\kappa a
-\int_0^a\mu_{nd}(\sigma)d\sigma\right)\int_0^a\frac{\alpha(\sigma)}{N^*}\exp\left(-\int_\sigma^a\beta(\varsigma)d\varsigma
-\lambda^*\int_\sigma^a\frac{\alpha(\varsigma)}{N^*}d\varsigma\right)d\sigma
da=1.\ee Substituting (\ref{4-11}) into ansatz (\ref{4-8}) and
taking into account (\ref{4-13}), (\ref{4-12}) and formulae
$\lambda^*=e^{-\kappa t}\lambda , \ N^*=e^{-\kappa t} N$, one
obtains the exact solution of the age-structured SIS
model~(\ref{4-7}).

Obviously age played an essentially   role during the COVID-19
pandemic because the deaths rate depends  very much on the age of
infected persons. Thus, development and  application of the
age-structured epidemic models in the modelling of the COVID-19
pandemic are important and several studies are devoted to these
aspects \cite{djilali-2020,bentout-2021,duan-2022}. In
\cite{djilali-2020}, the age-structured SEIR model \be\label{4-5}\ba
 \frac{\p s}{\p t}+\frac{\p s}{\p a} =-\alpha s \left(\int_0^{+\infty}i(a,t)da+\int_0^{+\infty}e(a,t)da\right), \medskip \\
  \frac{\p e}{\p t}+\frac{\p e}{\p a} =\alpha s \left(\int_0^{+\infty}i(a,t)da+\int_0^{+\infty}e(a,t)da\right)-\gamma e, \medskip \\
 \frac{\p i}{\p t}+\frac{\p i}{\p a} =\gamma e-\beta i -\mu(a)i, \medskip \\
 \frac{d r}{d t} =\beta\int_0^{+\infty}i(a,t)da,
 \ea\ee
 is considered in order to predict the epidemic peak outbreak  in South
Africa, Turkey and Brazil. In (\ref{4-5}), all parameters have the
same interpretations as  above. A generalization of the above model
(see (6) \cite{djilali-2020}) is also suggested by adding a new
subpopulation (compartment)  for individuals in quarantine.

 In
\cite{bentout-2021}, another approach was suggested by  considering
an age-structured model that takes into account two main components
of the COVID-19 pandemic: the number of infected individuals
requiring hospitalization, which leads to the estimation of required
beds, and the potential infection of healthcare personnel.
Consequently, the model predicts the timing of the peak and the
number of infectious cases at that peak both before and after the
implementation of nonpharmaceutical interventions. Additionally, a
comparison is made with the scenario of a full lockdown. In
\cite{duan-2022}, an age-structured model is proposed to study the
outbreak of the COVID-19 coronavirus in Wuhan, China. The value of
the basic reproduction number is computed in order to provide an
initial understanding of how contagious or virulent the pandemic  is
and how one might spread within a population.

 In order to take into
account both age and spatial heterogeneity, more complicated models
were developed \cite{kuniya-2015,kuniya-2018,kang-2021,tian-2022}.
In particular, the age-structured SIS epidemic model with  diffusion
\cite{kuniya-2015}
\begin{equation}\label{4-3}\begin{array}{l} \medskip  \frac{\p s}{\p t}+\frac{\p s}{\p a} = d_s \Delta s-\lambda s+\gamma i-\mu s,  \medskip\\
\frac{\p i}{\p t}+\frac{\p i}{\p a} = d_i \Delta i+\lambda s-\gamma
i-\mu i,
\end{array}\end{equation}
was further developed. In (\ref{4-3}), $s(a,t,x)$ and $i(a,t,x)$ are
the densities of susceptible and infective individuals of age $a
\geq 0$ at time $t$ in position $x \in \Omega \subset \mathbb{R}^n$,
respectively; $\mu(a, x)$ is the mortality of an individual of age
$a$ in position $x$; $\gamma(a,x)$ is the recovery rate of an
infective individual of age $a$ in position $x$; $\lambda(a,t, x)$
is the force of infection to a susceptible individual of age $a$ at
time $t$ in position $x$ given by formula
\[\lambda(a, t,
x)=\int_0^{a_{max}}\int_{\Omega}k(a,\sigma,x,y)i(\sigma,t,y)dyd\sigma,\]
where $k(a,\sigma,x,y)$ denotes the rate of disease transmission
from an infective individual of age $\sigma$ in position $y$ to a
susceptible individual of age $a$ in position $x$.

In \cite{kuniya-2015,kuniya-2018},  existence of  nontrivial
steady-states solutions of the  epidemic model (\ref{4-3}) is
investigated; the basic reproduction number $R_0$ for system in
question is estimated;  numerical simulations are performed to
verify the analytical results obtained.

In \cite{kang-2021},  well-posedness of an age-structured SIS model
is proved,  existence and uniqueness of the nontrivial steady state
corresponding to an endemic state are investigated, and the local
and global stability of this nontrivial steady state is studied.
Furthermore, the asymptotic properties of the principal eigenvalue
and the nontrivial steady state with respect to the nonlocal
diffusion rate are discussed.

In \cite{tian-2022},  the modified age-structured SIS model
\begin{equation}\nonumber\begin{array}{l} \medskip  \frac{\p s}{\p t} = d_s \Delta s-f(s)\int_0^{a_{max}}k(a)g(i(a,t,x))da,  \medskip\\
\frac{\p i}{\p t}+\frac{\p i}{\p a} = d_i \Delta i-\gamma i-\mu i,
\end{array}\end{equation}
 is studied. It can be noted that the function $s$, describing susceptible  individuals, does not depend on the variable $a$ (in contrast to the function $i$) but only on the maximal population age $a_{max}$.
  Here the smooth nonnegative
functions $f(s)$ and $g(i)$ satisfy  the conditions:
\[f(0)=g(0)=0, \ \frac{df}{ds}>0, \ \frac{dg}{di} \geq0, \ \frac{d^2g}{di^2}\leq0.\]
Moreover, the function $\frac{g(i)}{i}$ is continuously
differentiable and nonincreasing. Existence of travelling fronts of
the form $(s(a,t),\ i(a,t,x))= \left(S(x+\nu t),I(a,x+\nu t)\right)$
(here $\nu$ is the wave speed) of the above model is proved under
the following asymptotic boundary conditions \[S(-\infty)=S_0, \
S(+\infty)=S_1, \ I(a,-\infty)=I(a,+\infty)=0, \ 0\leq S_1\leq
S_0<\infty. \]
 The effects of nonlinear
functions $f$ and $g$
 and age structure on the basic reproduction number
and critical wave speed are  investigated in this paper as well.

 However, the basic  equations of the model,
multi-dimensional nonlinear integro-differential equations with
nonconstant coefficients, are very complicated. One needs to
simplify the model in order to derive results that are important
from applicability point of view.

\section{Conclusions}\label{sec-6}

 This review  provides a comprehensive analysis  of
mathematical models used for describing epidemic processes.  A main
focus is placed on the models that were developed and used for
modelling the COVID-19 pandemic. A huge number of studies was
published since the outbreak of COVID-19 in the end of 2019. So, it
is practically impossible to highlight even only  main results of
many hundreds of papers. Thus, we paid the main attention to the
studies devoted to the mathematical models based on {\it partial
differential equations} (typically reaction-diffusion equations) and
those involving not only numerical simulations but analytical
techniques as well.

The review  begins by acknowledging the historical significance of
the Kermack--McKendrick work, in which the SIR model was developed.
It is widely accepted  that  this model   laid  foundations for
mathematical modelling in epidemiology. In Section~\ref{sec-2}, we
also present  various extensions and  generalizations of the SIR
model such as the SIRD model, the  SEIR model, etc. All models of
this type are based on ODEs. Typically, numerical simulations have
been the primary methods for solving these models due to
nonlinearities in the basic ODEs. However, several  studies have
focused on finding exact and approximate
 solutions using relevant analytical  techniques
(Lie symmetry method, classical methods for integration of nonlinear
ODEs, the homotopy analysis method, optimal auxiliary functions
method, etc.). This review highlights importance of the results
obtained because exact solutions (even particular those)  are very
useful  for qualitative description of the pandemic spread and   for
estimating the accuracy of  numerical methods. Key epidemiological
parameters, like the basic reproduction number $R_0$,
 have been estimated using analytical results for these models.



During the COVID-19  pandemic, significant spatial heterogeneity was
observed in the pandemic spread, emphasizing the importance of
considering spatial aspects in mathematical modelling. It is
well-known that reaction-diffusion equations are  applied  for
mathematical description  of a wide range of biomedical processes in
order to take into account the spatial dynamics.
Recently, these equations were used  to study the spatial spread of
the COVID-19 pandemic. The main part of the review,
Sections~\ref{sec-3} and~\ref{sec-4}, is devoted to such type models
because they are the ones that allow us to better understand the
spatial dynamics of epidemics.

We start from the natural generalizations of the classical models,
in particular, the SIR and SEIR models with diffusion in space.
Conditions for the existence of travelling waves, relation with the
diffusive Lotka--Volterra system, estimation of the basic
reproduction number are discussed. More complicated models,
especially those based on multi-component systems and/or involving
nonconstant diffusivities are presented as well and their
application for numerical estimations  of  the spatial spread of the
COVID-19 pandemic is discussed.
 Furthermore, epidemic models  involving reaction-diffusion
equations with cross-diffusion and convective terms are discussed.
Cross-diffusion is a known phenomenon observed in various biomedical
and ecological  processes. Recently,   reaction-diffusion systems
with cross-diffusion were  applied to the mathematical modelling of
the COVID-19 spread. In particular, exact solutions in the form of
travelling waves were constructed and their interpretation provided.

Finally, the review touches upon age-structured epidemic models.
Such type models   incorporate age dimension  as a second time
variable. Usually relevant  governing equations are
integro-differential, however, their plausible approximations can be
derived in the form of PDEs. In this case,  systems of the
first-order PDEs are obtained and exact solutions (at least in
implicit forms) can be derived. However, if one needs to take into
account also spatial heterogeneity then relevant models are much
more complicated. Recently, some analytical results, especially
conditions for existence of travelling fronts, were derived.
However, to the best of our knowledge, there are no applications of
such models for numerical simulation of  the spatial spread of the
COVID-19 pandemic.

It should be noted that there are some papers, see, e.g.
\cite{mac-2021,zhou-2021}, in which the models based on
reaction-diffusion equations  with time delays are suggested for
describing epidemic processes. Actually, such type  models are
natural extensions of the models based on ODEs with time delays
(see, e.g., Chapter 10 in \cite{brauer-12}).  However, to the best
of our knowledge, there are no direct applications for modelling the
COVID-19 spread in those papers.

 In summary, this review could  serve as a valuable
resource for researchers and practitioners in the field of modelling
in epidemiology, offering analysis and application of a wide range
of mathematical models. The main emphasis is placed on the models
taking into account  the spatial heterogeneity that was widely
observed during the COVID-19 pandemic spread.
  We  presented  a comprehensive
overview of the studies, especially those published after the
outbreak of COVID-19, devoted  to
 epidemic models and  based on partial differential equations.
 \medskip

\textbf{Acknowledgments:} V.~Davydovych and V.~Dutka acknowledge
that this research was supported by the National Research Foundation
of Ukraine, project 2021.01/0311. R.Ch. acknowledges that this
research was partly  funded by the British Academy's Researchers at
Risk Fellowships Programme.


\begin{thebibliography}{99}

\bibitem {brauer-12} Brauer,  F.;  Castillo-Chavez, C.
\emph{Mathematical Models in Population Biology and Epidemiology};
Springer:  New York, 2012.

\bibitem {diekmann-2000} Diekmann, O.; Heesterbeek, J.A.P. \emph{Mathematical Epidemiology of
Infectious Diseases: Model Building, Analysis and Interpretation};
John Wiley \& Sons: Chichester, UK, 2000.

\bibitem {keeling-2008}  Keeling, M.J.;   Rohani, P.  \emph{Modeling Infectious Diseases in Humans
and Animals};  Princeton University Press:  Princeton, USA, 2008.

\bibitem {murray-1989}  Murray, J.D. \emph{Mathematical Biology};  Springer: Berlin, 1989.

 \bibitem {murray-2003}  Murray, J.D.
\emph{Mathematical Biology, II: Spatial
 Models and Biomedical Applications}; Springer: Berlin, 2003.

 \bibitem {hadeler-2017}  Hadeler K.P. \emph{Topics in Mathematical
Biology}; Berlin: Springer, 2017.

\bibitem {bailey-1975} Bailey, N.T.J.
\emph{ The Mathematical Theory of Infectious Diseases and Its
Applications.}
 Charles Griffin \& Company: London, 1975.

\bibitem{kermack-1927}  Kermack, W.O.;  McKendrick, A.G. A contribution to the mathematical
theory of epidemics. \emph{Proc. Roy. Soc. A} \textbf{1927},
\emph{115}, 700--721.

\bibitem {dietz}  Dietz,  K.  \emph{ The  Incidence  of  Infectious  Diseases Under  the  Influence  of  Seasonal
Fluctuations};  Lecture  Notes  in  Biomathematics  \emph{11},
Springer: Berlin,  1976, pp. 1--15.

\bibitem {anderson} Anderson,  R.M.; May,   R.M.    Directly  transmitted  infectious diseases:  Control  by
vaccination. \emph{Science} \textbf{1982},  \emph{215}, 1053--1060.

\bibitem{kermack-1932} Kermack, W. O., McKendrick, A. G.  Contributions to the
mathematical theory of epidemics. II. -- The problem of endemicity.
\emph{Proc. Roy. Soc. A} \textbf{1932}, \emph{138}, 55--83.

\bibitem {lin-2010} Lin, F.; Muthuraman, K.;  Lawley, M. An optimal control
theory approach to non-pharmaceutical interventions. \emph{BMC
Infectious Diseases} \textbf{2010}, \emph{10}, 1--13.


\bibitem{yang-2021} Yang, W.; Zhang, D.; Peng, L.; Zhuge, C.;  Hong, L.
Rational evaluation of various epidemic models based on the COVID-19
data of China. \emph{Epidemics} \textbf{2021}, \emph{37}, 100501.

\bibitem{ch-dav-preprint20} Cherniha, R.; Davydovych, V. A mathematical model for the COVID-19
outbreak.  \emph{ArXiv} \textbf{2020},  arXiv:2004.01487v2.

 \bibitem{ch-dav-2020} Cherniha, R.; Davydovych, V. A mathematical model for the COVID-19
  outbreak and its applications. \emph{Symmetry} \textbf{2020},  \emph{12}, 12 pp.



\bibitem{nesteruk-2021} Nesteruk I. \emph{COVID19 Pandemic Dynamics}.
Springer Nature:  Singapore, 2021.

\bibitem{fanelli-2020} Fanelli, D.; Piazza, F. Analysis and forecast of COVID-19
spreading in China, Italy and France. \emph{Chaos, Solitons \&
Fractals} \textbf{2020}, \emph{134}, 109761.


\bibitem{alenezi-2021} Alenezi, M.N.; Al-Anzi, F.S.; Alabdulrazzaq, H.; Alhusaini,
A.; Al-Anzi, A.F.  A study on the efficiency of the estimation
models of COVID-19. \emph{Results in Physics} \textbf{2021},
\emph{26}, 104370.

\bibitem{jai-2022} El Jai, M.; Zhar, M.; Ouazar, D.; Akhrif, I.;  Saidou, N.
Socio-economic analysis of short-term trends of COVID-19: modelling
and data analytics. \emph{BMC Public Health} \textbf{2022},
\emph{22}, 1633.

\bibitem{kalachev-2023} Kalachev, L.; Landguth, E. L.;  Graham, J. Revisiting
classical SIR modelling in light of the COVID-19 pandemic.
\emph{Infectious Disease Modelling} \textbf{2023}, \emph{8}, 72--83.

\bibitem{kyrychko-2020} Kyrychko, Y.N.; Blyuss, K.B.;  Brovchenko, I.
Mathematical modelling of the dynamics and containment of COVID-19
in Ukraine. \emph{Scientific reports} \textbf{2020}, \emph{10},
19662.


\bibitem{alimohamadi-2020} Alimohamadi, Y.; Taghdir, M.; Sepandi, M.  Estimate of the
basic reproduction number for COVID-19: a systematic review and
meta-analysis. \emph{J. Prev. Med. Public. Health} \textbf{2020},
\emph{53}, 151.

\bibitem{salom-2021} Salom, I.; Rodic, A.; Milicevic, O.; Zigic, D.; Djordjevic,
M.; Djordjevic, M.  Effects of demographic and weather parameters on
COVID-19 basic reproduction number. \emph{Front. Ecol. Environ.}
\textbf{2021}, \emph{8}, 617841.

\bibitem{barlow-2020} Barlow, N.S.;  Weinstein, S.J.  Accurate closed-form
solution of the SIR epidemic model. Physica D \textbf{2020},
\emph{408}, 132540.

\bibitem{chen-2021} Chen, X.; Li, J.; Xiao, C.;  Yang, P. Numerical solution
and parameter estimation for uncertain SIR model with application to
COVID-19. \emph{Fuzzy Optim. Decis. Making}  \textbf{2021},
\emph{20}, 189-208.

\bibitem{zhu-2021} Zhu, X.; Gao, B.; Zhong, Y.; Gu, C.;  Choi, K. S. Extended
Kalman filter based on stochastic epidemiological model for COVID-19
modelling.  \emph{Comput. Biol. Med.} \textbf{2021}, \emph{137},
104810.


\bibitem{nucci-2004} Nucci, M.C.; Leach, P.G.L. An integrable SIS model. \emph{ J. Math. Anal. Appl.} \textbf{2004}, \emph{290}, 506--518.


\bibitem{harko-2014}  Harko, T.; Lobo,F.S.N.;  Mak, M. Exact analytical solutions of
the Susceptible-Infected-Recovered (SIR) epidemic model and of the
SIR model with equal death and birth rates. \emph{Appl. Math.
Comput.} \textbf{2014}, \emph{236}, 184--194.



\bibitem{yoshida-2022} Yoshida, N. Exact solution of the
Susceptible-Infectious-Recovered-Deceased (SIRD) epidemic model.
\emph{Electron. J. Qual. Theory Differ. Equ.} \textbf{2022},
\emph{38}, 1--24.


\bibitem{yoshida-2023} Yoshida, N. Existence of exact solution of the
Susceptible-Exposed-Infectious-Recovered (SEIR) epidemic model.
\emph{J. Diff. Equ.}  \textbf{2023}, \emph{355}, 103--143.



\bibitem{khan-2009} Khan, H.; Mohapatra, R.N.; Vajravelu, K.;  Liao, S.J.  The
explicit series solution of SIR and SIS epidemic models. \emph{Appl.
Math. Comput.}  \textbf{2009}, \emph{215}, 653--669.

\bibitem{marinca-2021} Marinca, B.; Marinca, V.;  Bogdan, C. Dynamics of SEIR
epidemic model by optimal auxiliary functions method. \emph{Chaos,
Solitons \& Fractals} \textbf{2021}, \emph{147}, 110949.


\bibitem{kendall-1965}  Kendall, D.G. Mathematical models of the spread of
infection. \emph{Mathematics and Computer Science in Biology and
Medicine} \textbf{1965}, 213--225.

\bibitem{radcliffe-1973} Radcliffe, J. The initial geographical spread of host-vector and
carrier-borne epidemics. \emph{J. Appl. Prob.} \textbf{1973},
\emph{10}, 703--717.

\bibitem{noble-1974} Noble, J.V. Geographic and temporal development of plagues. \emph{Nature}
\textbf{1974}, \emph{250}, 726--729.

\bibitem{murray-1985} Kall\'{e}n, A.;  Arcuri, P.; Murray, J.D.  A simple model for the
spatial spread and control of rabies. \emph{J. Theor. Biol.}
\textbf{1985}, \emph{116}, 377--393.

\bibitem{kallen-1984} Kall\'{e}n, A. Thresholds and travelling waves in an epidemic model
 for rabies.\emph{ Nonlinear Anal.} \textbf{1984}, \emph{8}, 851--856.



\bibitem{zhang-2020} Zhang, L.; Wang, Z.C.; Zhao, X.Q. Time periodic
traveling wave solutions for a Kermack--McKendrick epidemic model
with diffusion and seasonality. \emph{J. Evol. Equ.} \textbf{2020},
\emph{20}, 1029--1059.





\bibitem{cheng-2022}  Cheng, Z.; Wang, J.  Modeling epidemic
flow with fluid dynamics. \emph{Math. Biosci. Eng.} \textbf{2022},
\emph{19}, 8334--8360.

\bibitem{zhi-2023} Zhi, S.; Niu, Y.T.; Su, Y.H.; Han, X.  Influence of human behavior on COVID-19 dynamics based on a reaction-diffusion model.
 \emph{Qual. Theory Dyn. Syst.} \textbf{2023}, \emph{22}, 26 pp.

\bibitem{viguerie-2021}    Viguerie, A.; Lorenzo, G.; Auricchio, F. Simulating the spread of COVID-19 via a spatially-resolved
susceptible-exposed-infected-recovered-deceased (SEIRD) model with
heterogeneous diffusion. \emph{Appl. Math. Lett.} \textbf{2021},
\emph{111}, 106617.

\bibitem{viguerie-2020} Viguerie, A.; Veneziani, A.; Lorenzo, G. Diffusion-reaction models
in a continuum mechanics framework with application to COVID-19
modelling. \emph{Comput. Mech.} \textbf{2020}, \emph{66},
1131--1152.



\bibitem{grave-2021} Grave, M.;  Coutinho, A.L. Adaptive mesh refinement and
coarsening for diffusion-reaction epidemiological models.
\emph{Comput. Mech.}  \textbf{2021}, \emph{67}, 1177--1199.


\bibitem{grave-2022} Grave, M.; Viguerie, A.; Barros, G.F.; Reali, A.; Andrade Roberto, F.S.; Coutinho Alvaro, L.G.A.
 Modeling nonlocal behavior in epidemics via a
 reaction-diffusion system incorporating population movement
 along a network. Comput. Methods Appl. Mech. Engrg. \textbf{2022},
\emph{401}, 115541.


\bibitem{zhu-2019} Zhu, C.C.; Zhu, J.; Liu, X.L.
 Influence of spatial heterogeneous environment on long-term
 dynamics of a reaction-diffusion
 SVIR epidemic model with relapse. \emph{Math. Biosci. Eng.}\textbf{ 2019}, \emph{16}, 5897--5922.

\bibitem{tu-2023} Tu, Y.; Hayat, T.; Hobiny, A.; Meng, X. Modeling and
multi-objective optimal control of reaction-diffusion COVID-19
system due to vaccination and patient isolation. \emph{Appl. Math.
Model.} \textbf{2023}, \emph{118}, 556--591.


\bibitem{mammeri-2020} Mammeri, Y.  A reaction-diffusion system to better comprehend the unlockdown: Application of SEIR-type model
 with diffusion to the spatial spread of COVID-19 in France. \emph{Comput. Math. Biophys.} \textbf{2020}, \emph{8}, 102--113.

\bibitem{yin-2022}  Yin, H.M. On a reaction-diffusion system modelling infectious diseases without lifetime immunity.
 \emph{Euro. J. Appl. Math.} \textbf{2022}, \emph{33}, 803--827.







\bibitem{capasso-1994} Capasso, V.;  Di Liddo, A. Asymptotic behaviour of
reaction-diffusion systems in population and epidemic models: the
role of cross diffusion. \emph{J. Math. Biol.}  \textbf{1994},
\emph{32}, 453--463.

\bibitem{bendahmane-2010} Bendahmane, M.; Langlais, M.  A reaction-diffusion
system with cross-diffusion modelling the spread of an epidemic
disease. \emph{J. Evol. Equ.} \textbf{2010}, \emph{10}, 883--904.

\bibitem{ch-dav-2022-cd} Cherniha, R.M.;   Davydovych, V.V. A reaction-diffusion
system with cross-diffusion: Lie symmetry, exact solutions and their
applications in the pandemic modelling. \emph{Euro. J. Appl. Math.}
\textbf{2022}, \emph{33}, 785--802.



\bibitem {kel-seg-71}   Keller, E.K.;  Segel, L.A. Traveling bands of chemotactic bacteria:
A Theoretical Analysis. \emph{J. Theor. Biol.} \textbf{1971},
\emph{30}, 235--248.


\bibitem {sh-ka-te}  Shigesada, N.;  Kawasaki, K.;  Teramoto, E.
Spatial segregation of interacting species. \emph{J. Theoret. Biol.}
\textbf{1979}, \emph{79}, 83--99.

\bibitem {ch-da-ki-23} Cherniha, R.; Davydovych, V.; King, J.R. The  Shigesada--Kawasaki--Teramoto  model:
conditional symmetries, exact solutions and their properties.
\emph{Comm. Nonlinear Sci. Numer. Simulat.} \textbf{2023},
\emph{124}, 107313;


\bibitem{kuniya-2015} Kuniya, T.;  Oizumi, R. Existence result for an
age-structured SIS epidemic model with spatial diffusion.
\emph{Nonlinear Anal.: Real World Appl.} \textbf{2015}, \emph{23},
196--208.

\bibitem{kuniya-2018} Kuniya, T.;  Inaba, H.;  Yang, J. Global behavior of SIS epidemic
models with age structure and spatial heterogeneity. \emph{Jpn J.
Ind. Appl. Math.} \textbf{2018}, \emph{35}, 669--706.

\bibitem{kang-2021} Kang, H.; Ruan, S. Mathematical analysis on an
age-structured SIS epidemic model with nonlocal diffusion. \emph{J.
Math. Biol.} \textbf{2021}, \emph{83}, 5.


\bibitem{tian-2022} Tian, X.;  Guo, S. Traveling waves of an epidemic model
with general nonlinear incidence rate and infection-age structure.
\emph{Z. Angew. Math. Phys.} \textbf{2022}, \emph{73}, 167.

\bibitem{loli-2020} Loli Piccolomini, E.;  Zama, F. Monitoring Italian COVID-19
spread by a forced SEIRD model. \emph{PloS One} \textbf{2020},
\emph{15}, e0237417.

\bibitem {verh-1838}  Verhulst,  P.F. Notice sur la loi que la
population suit dans son accroissement. \emph{Corr. Math. Physics.}
\textbf{1838}, \emph{10}, 113.

\bibitem {meters} https://www.worldometers.info/coronavirus


 \bibitem{ch-dav-book}  Cherniha, R.;  Davydovych, V.
\emph{Nonlinear Reaction-Diffusion Systems --- Conditional Symmetry,
exact Solutions and Their Applications in Biology};  Lecture Notes
in Mathematics  \emph{2196}. Springer: Cham, 2017.

 \bibitem {ch-dav-2022} Cherniha, R.; Davydovych, V.  Construction and application of exact
   solutions of the diffusive Lotka--Volterra system: A review and new results.
   \emph{Comm. Nonlinear Sci. Numer. Simulat.} \textbf{2022}, \emph{113}, 106579.

\bibitem {ahmed-2021} Ahmed, N.; Elsonbaty, A.; Raza, A.; Rafiq, M.; Adel, W. Numerical
simulation and stability analysis of a novel reaction-diffusion
COVID-19 model. \emph{Nonlinear Dyn.}, \textbf{2021}, 106,
1293--1310.

\bibitem {keller-2013} Keller, J.P.; Gerardo-Giorda, L.;  Veneziani, A. Numerical
simulation of a susceptible-exposed-infectious space-continuous
model for the spread of rabies in raccoons across a realistic
landscape. \emph{J. Biol. Dyn.} \textbf{2013}, \emph{7}, 31--46.

\bibitem {Zhuang-2021} Zhuang, Q.;  Wang, J. A
spatial epidemic model with a moving boundary.
 \emph{Infect. Dis. Model.} \textbf{2021}, \textbf{6}, 1046--1060.


\bibitem {greenhalgh-1987} Greenhalgh, D. Analytical results on the stability of
age-structured recurrent epidemic models. \emph{IMA J Math. Appl.
Med. Biol.} \textbf{1987}, \emph{4}, 109--144.

\bibitem {busenberg-1988} Busenberg, S.; Cooke, K.;  Iannelli, M. Endemic thresholds
and stability in a class of age-structured epidemics. \emph{SIAM J.
Appl. Math.} \textbf{1988}, \emph{48}, 1379--1395.

\bibitem {inaba-1990} Inaba, H. Threshold and stability results for an
age-structured epidemic model. \emph{J. Math. Biol.} \textbf{1990},
\emph{28}, 411--434.


\bibitem {djilali-2020} Djilali, S.;  Ghanbari, B.  Coronavirus pandemic: A predictive analysis of the peak
outbreak epidemic in South Africa, Turkey, and Brazil. \emph{Chaos,
Solitons \& Fractals} \textbf{2020}, \emph{138}, 109971.

\bibitem {bentout-2021} Bentout, S.; Tridane, A.; Djilali, S.;  Touaoula, T.M.  Age-structured modelling of COVID-19
epidemic in the USA, UAE and Algeria. \emph{Alex. Eng. J. }
\textbf{2021}, \emph{60}, 401--411.

\bibitem {duan-2022} Duan, X.C.; Li, X.Z.; Martcheva, M.;  Yuan, S. Using an
age-structured COVID-19 epidemic model and data to model virulence
evolution in Wuhan, China. \emph{J. Biol. Dyn.} \textbf{2022},
\emph{16}, 14--28.

\bibitem {mac-2021}  Macias-Diaz, J. E.; Ahmed, N.;
 Jawaz, M.; Rafiq, M.; Aziz ur Rehman, M. Design and analysis of a discrete method for a time-delayed
 reaction-diffusion epidemic model. \emph{Math. Methods Appl. Sci.}   \textbf{2021}, \emph{44}, 5110--5122.

\bibitem {zhou-2021}  Zhou, J.; Ma, X.; Yang, Y.; Zhang, T.
 A diffusive SVEIR epidemic model with time delay and general incidence. \emph{Acta. Math. Sci.} \textbf{2021}, \emph{41}, 1385--1404.


\end{thebibliography}
\end{document}